\documentclass[twocolumn]{aastex631}
\usepackage[varg]{txfonts}
\usepackage{graphicx}
\usepackage{graphics}
\usepackage{amssymb}
\usepackage{amsmath}
\usepackage[T1]{fontenc}
\usepackage[utf8]{inputenc}
\usepackage{color} 
\newcommand{\g}{\ensuremath{\gamma}}
\newcommand{\p}{^{\prime}}

\newcommand{\E}[1]{\times 10^{#1}}

\newcommand{\source}{\object{PKS~1510$-$089}}
\newcommand{\hess}{H.E.S.S.}
\newcommand{\fermi}{\textit{Fermi}-LAT}
\newcommand{\stat}{_{\rm stat}}
\newcommand{\sys}{_{\rm sys}}
\received{}
\revised{}
\accepted{}
\submitjournal{ApJL}
\shorttitle{The vanishing of the primary emission region in \source}
\shortauthors{H.E.S.S.~Collaboration~et~al.}
\begin{document}
\title{The vanishing of the primary emission region in \source}
\correspondingauthor{J.~Barnard, M.~B\"ottcher, H.M.~Schutte, M.~Zacharias, Email:~contact.hess@hess-experiment.eu
}

\author{F.~Aharonian}
\affiliation{Dublin Institute for Advanced Studies, 31 Fitzwilliam Place, Dublin 2, Ireland}
\affiliation{Max-Planck-Institut f\"ur Kernphysik, P.O. Box 103980, D 69029 Heidelberg, Germany}
\affiliation{Yerevan State University,  1 Alek Manukyan St, Yerevan 0025, Armenia}

\author{F.~Ait~Benkhali}
\affiliation{Landessternwarte, Universit\"at Heidelberg, K\"onigstuhl 12, D 69117 Heidelberg, Germany}

\author{J.~Aschersleben}
\affiliation{Kapteyn Astronomical Institute, University of Groningen, Landleven 12, 9747 AD Groningen, The Netherlands}

\author[0000-0002-2153-1818]{H.~Ashkar}
\affiliation{Laboratoire Leprince-Ringuet, École Polytechnique, CNRS, Institut Polytechnique de Paris, F-91128 Palaiseau, France}

\author[0000-0002-9326-6400]{M.~Backes}
\affiliation{University of Namibia, Department of Physics, Private Bag 13301, Windhoek 10005, Namibia}
\affiliation{Centre for Space Research, North-West University, Potchefstroom 2520, South Africa}

\author[0000-0002-5085-8828]{V.~Barbosa~Martins}
\affiliation{DESY, D-15738 Zeuthen, Germany}

\author{J.~Barnard}
\affiliation{Department of Physics, University of the Free State,  PO Box 339, Bloemfontein 9300, South Africa}

\author[0000-0002-5797-3386]{R.~Batzofin}
\affiliation{Institut f\"ur Physik und Astronomie, Universit\"at Potsdam,  Karl-Liebknecht-Strasse 24/25, D 14476 Potsdam, Germany}

\author[0000-0002-2115-2930]{Y.~Becherini}
\affiliation{Université de Paris, CNRS, Astroparticule et Cosmologie, F-75013 Paris, France}
\affiliation{Department of Physics and Electrical Engineering, Linnaeus University,  351 95 V\"axj\"o, Sweden}

\author[0000-0002-2918-1824]{D.~Berge}
\affiliation{DESY, D-15738 Zeuthen, Germany}
\affiliation{Institut f\"ur Physik, Humboldt-Universit\"at zu Berlin, Newtonstr. 15, D 12489 Berlin, Germany}

% ...

\collaboration{153}{H.E.S.S. Collaboration}


\author{D.A.H.~Buckley} 
\affiliation{South African Astronomical Observatory, PO Box 9, Observatory 7935, South Africa}
\affiliation{Southern African Large Telescope Foundation, PO Box 9, Observatory 7935, South Africa}
\affiliation{Department of Physics, University of the Free State, PO Box 339, Bloemfontein 9300, South Africa}
\affiliation{Department of Astronomy, University of Cape Town, Private Bag X3, Rondebosch 7701, South Africa}

\author{J.~Cooper} 
\affiliation{Department of Physics, University of the Free State,  PO Box 339, Bloemfontein 9300, South Africa}

\author{D.~Groenewald} 
\affiliation{South African Astronomical Observatory, PO Box 9, Observatory 7935, South Africa}
\affiliation{Southern African Large Telescope Foundation, PO Box 9, Observatory 7935, South Africa}
\nocollaboration{3}
%
%
\begin{abstract}
In July 2021, \source\ exhibited a significant flux drop in the high-energy \g-ray (by a factor $10$) and optical (by a factor $5$) bands and remained in this low state throughout 2022. Similarly, the optical polarization in the source vanished, resulting in the optical spectrum being fully explained through the steady flux of the accretion disk and the broad-line region. Unlike the aforementioned bands, the very-high-energy \g-ray and X-ray fluxes did not exhibit a significant flux drop from year to year. This suggests that the steady-state very-high-energy \g-ray and X-ray fluxes originate from a different emission region than the vanished parts of the high-energy \g-ray and optical jet fluxes. The latter component has disappeared through either a swing of the jet away from the line-of-sight or a significant drop in the photon production efficiency of the jet close to the black hole. Either change could become visible in high-resolution radio images.
\end{abstract}
%
%
%
%
%
%
%
\section{Introduction}
%

As the relativistic jets of blazars are almost aligned with the line-of-sight, the emission region producing most of the jet's radiation can be studied in great detail owing to the Doppler beaming of the radiation. The observed variability implies a compact emission region leading to the one-zone model \citep[e.g.,][]{boettcher19}. In the leptonic version of this model, a single electron distribution is responsible for the multiwavelength (MWL) emission through synchrotron emission and inverse-Compton (IC) scattering of ambient photon fields, such as synchrotron, accretion disk (AD), broad-line region (BLR) or dusty torus (DT) photons. In some extensions of the model, relativistic protons may also influence the production of \g\ rays \citep[for more details on the radiation processes, see e.g.,][]{boettcherHarrisKrawczynski12,cerruti20}.

\source\ is a flat-spectrum radio quasar (FSRQ) at redshift $z=0.361$ \citep{burbidgekinman66}. It is one of the few FSRQs detected at very-high-energy (VHE, $E>100\,$GeV) \g\ rays\footnote{For an up-to-date list, see \url{http://tevcat2.uchicago.edu/}.} \citep{hess13}. FSRQs are blazars with bright optical emission lines implying the presence of a strong BLR. Hence, the VHE emission zone must be located at the edge of or beyond the BLR in order to avoid the strong absorption of VHE photons. In turn, models were developed that explained the spectral energy distribution (SED) of \source\ either through the necessity of multiple target photon fields for the IC process \citep[e.g.,][]{barnacka+14} or through two spatially separate emission zones \citep{nalewajko+12,prince+19} with a primary emission zone within the BLR and a secondary emission zone several parsec from the black hole within the DT.
\source\ is known for its complex MWL behavior \citep[e.g.,][]{brown13,saito+15,zacharias+19} without clear correlation patterns between energy bands. One of the most spectacular flares was the VHE flare in 2016 \citep{hess+21} with only moderate counterparts in the high-energy (HE, $E>100\,$MeV) \g-ray and optical bands.

However, unlike all other FSRQs detected at VHE \g\ rays, \source\ also emits VHE photons in times of quiescence. \cite{magic18} integrated their data taken during times without any MWL flaring activity. Their VHE spectrum is a near-perfect continuation of the HE spectrum allowing for the application of the one-zone model in both a near-zone and a far-zone scenario. In the near-zone scenario, the emission region is located close to the edge of the BLR about $0.1\,$pc from the black hole, while the far-zone emission region is located at about 1\,pc from the black hole within the DT. Similarly, \cite{Meyer+19} independently derived a HE \g-ray low-state spectrum of \source, which they coupled with radio and X-ray observations of the extended kpc-scale jet explaining the SED in terms of an IC model scattering the cosmic microwave background (CMB).

In this paper, a sudden change in the appearance of \source\ is reported. While flares had become less and less frequent since about 2017,\footnote{See, e.g., the public \fermi\ light curves: \url{https://fermi.gsfc.nasa.gov/ssc/data/access/lat/msl\_lc/source/1510-089}.}
in July 2021 the source suddenly and abruptly dropped in HE and optical flux as seen in observations with \fermi\ and ATOM, respectively. Similarly, the optical polarization in \source\ measured with SALT vanished. Meanwhile, the VHE and X-ray fluxes observed with H.E.S.S. and the \textit{Neil Gehrels Swift} observatory (hereafter \textit{Swift}), respectively, remained almost steady. 

%
%
\section{Data analysis} \label{sec:ana}
%

%
\subsection{Very-high-energy \g\ rays} \label{sec:hess}
The five telescopes of the \hess\ array recording VHE \g\ rays are located in the Khomas Highland in Namibia at an altitude of about $1800\,$m. Four telescopes (CT1-4) with 106\,m$^2$ mirror area each, are laid out in a square of 120\,m side length giving an optimal energy threshold of $\sim 100\,$GeV. A fifth telescope (CT5) with 600\,m$^2$ mirror area is located in the center of the square. 
In this study, data recorded with CT1-4 are used. 

For the observations in 2021 (MJD 59311-59382) and 2022 (MJD 59672-59794), standard quality selection \citep{hess06} results in acceptance corrected observation times of $50.9\,$h in 2021 and $36.5\,$h in 2022, respectively. The data sets have been analyzed with the Model analysis chain \citep{denauroisrolland09} using \textsc{very loose} cuts. These cuts provide the lowest possible energy threshold with $129\,$GeV and $106\,$GeV in 2021 and 2022, respectively. The results have been cross-checked and verified using the independent reconstruction and analysis chain ImPACT \citep{parsonshinton14} providing consistent results. 
\source\ is detected with a significance of $13.5\sigma$ in 2021, and with $10.3\sigma$ in 2022.

In order to derive the light curves and photon spectra, instrument response functions were created using \texttt{Run Wise Simulations} \citep{holler20}, which accurately reproduce the atmospheric and instrumental conditions for each observation. There is no significant variability in the period-wise light curve [cf., Fig.~\ref{fig:mwl_lightcurve}(a)].

In both years, the spectra are consistent with power laws of the form

\begin{align}
 F(E) = N(E_0)\times\left( \frac{E}{E_0} \right)^{-\Gamma} 
 \label{eq:pwl},
\end{align}
where $N$ is the normalization at decorrelation energy $E_0$, and $\Gamma$ is the spectral index. The parameters for 2021 are $N=(17\pm 1\stat{}^{+6}_{-5}{}\sys)\E{-12}\,$ph\,cm$^{-2}$s$^{-1}$TeV$^{-1}$, $E_0=256\,$GeV, and $\Gamma=3.4\pm 0.1\stat\pm 0.4\sys$. In 2022, the spectral parameters are $N=(8.8\pm 0.7\stat{}^{+2.9}_{-2.4}{}\sys)\E{-12}\,$ph\,cm$^{-2}$s$^{-1}$TeV$^{-1}$, $E_0=296\,$GeV, and $\Gamma=3.0\pm 0.1\stat\pm 0.4\sys$. The main systematic error is the uncertainty of $10\%$ on the energy scale. 

The spectra are shown in Fig.~\ref{fig:spectra} (top) along with spectra from the detection \citep{hess13} and the low-state spectrum of \citep{magic18}. The latter is compatible with both spectra of 2021 and 2022, while the initial detection spectrum agrees with the new ones at the highest energies.

%
\subsection{High-energy \g\ rays} \label{sec:fermi}
%
\fermi\ monitors the HE $\gamma$-ray sky every three hours in the energy range from $20\,$MeV to beyond $300\,$GeV \citep{atwood09}. 
The analysis was performed with the \textsc{FermiTools}\footnote{\url{https://github.com/fermi-lat/Fermitools-conda/wiki}} version 2.2.0 software package employing the \textsc{P8R3\_SOURCE\_V3}\footnote{\url{http://fermi.gsfc.nasa.gov/ssc/data/analysis/documentation/Cicerone/Cicerone\_LAT\_IRFs/IRF\_overview.html}} 
instrument response functions and the \textsc{gll\_iem\_v07} and \textsc{iso\_P8R3\_SOURCE\_V3\_v1} models\footnote{\url{http://fermi.gsfc.nasa.gov/ssc/data/access/lat/BackgroundModels.html}} for the Galactic and isotropic diffuse emissions \citep{acero16}, respectively.
A binned analysis of the \textsc{SOURCE} class events between energies of $100\,$MeV and $500\,$GeV was performed for a region of interest (ROI) with radius $10^{\circ}$ centred at the nominal position of \source. In order to reduce contamination from the Earth Limb, a zenith angle cut of $90^{\circ}$ was applied. 
Sources within a region of radius $15^{\circ}$ around \source\ listed in the 4FGL-DR3 catalog \citep{abdollahi20,ajello20} have been accounted for in the likelihood analysis. 

The likelihood fitting procedure is iterative \citep[for more details, see][Section 3.1]{2018A+C....22....9L}.
First, all parameters from a source are fixed if a hint of emission from that object is detected with a test statistics\footnote{The TS value is defined as twice the difference of log-likelihood values of the optimised ROI model with and without the source included, $\mathrm{TS} = -2(\ln\mathcal{L}_1 - \ln\mathcal{L}_0)$~\citep{mattox96}.} of $\mathrm{TS}<9$ and if the predicted number of photons from that source contributes less than 5\% of the total of photon counts within the ROI. 
Second, only spectral parameters of sources within $3^\circ$ from \source\ are left free to vary.
All other source parameters are fixed to their respective 4FGL values, which are also used for all sources included in the model as seed inputs. 
The normalization of the Galactic and isotropic background templates are left as additional free parameters. 
Neither the residual nor count maps show any particular hot spots above a significance at the $\sim2\,\sigma$ level. Therefore, the best-fit model describes the ROI well.

The best-fit ROI model is then used to derive light curves of \source\ in the time range from January 2021 to September 2022 with a binning of 3 and 7 days, respectively. They are shown in Fig.~\ref{fig:mwl_lightcurve}(b). In the first half of 2021, the light curve was variable within a factor of 3 around its average integral flux of $\sim 4.3\E{-7}\,$ph\,cm$^{-2}$s$^{-1}$ in the $[100\,$MeV$; 500\,$GeV$]$ energy range. This average is below the 4FGL-DR3 catalog [indicated by the gray dashed line in Fig.~\ref{fig:mwl_lightcurve}(b)]. However, on 2021 July 18 (MJD~59413) the flux decreased significantly to an average value of $\sim 6\E{-8}\,$ph\,cm$^{-2}$s$^{-1}$, which is more than one order of magnitude below the 4FGL-DR3 value.

For the spectral analysis, two time ranges have been considered that coincide with the H.E.S.S. observation windows in 2021 (MJD 59311-59382) and 2022 (MJD 59672-59794).
In 2021, the differential photon spectrum of \source\ is described with a log-parabola function, which improves the spectral fit with respect to a pure power-law at a $3.3\sigma$ confidence level,

\begin{align}
 \frac{dN}{dE} = N(E_0)\times\left( \frac{E}{E_0} \right)^{-\Gamma-\beta\log{(E/E_0)}} 
 \label{eq:logp},
\end{align}
with normalization $N=(3.61 \pm 0.31\stat)\E{-11}\,$ph\,cm$^{-2}$s$^{-1}$MeV$^{-1}$, pivot energy $E_0=881\,\text{MeV}$ fixed at the 4FGL-DR3 value, photon index $\Gamma = 2.42\pm 0.07\stat$ and curvature $\beta = 0.05\pm 0.04\stat$. This spectrum is fully compatible with the 4FGL-DR3 catalog except for the normalization. 
In 2022, the spectrum is compatible with a simple power-law\footnote{A log-parabolic spectral shape is also tested for, but does not yield a better fit of the data with respect to a power-law.} with normalization $N=(7.36 \pm 0.92\stat)\E{-12}\,$ph\,cm$^{-2}$s$^{-1}$MeV$^{-1}$, pivot energy $E_0=881\,\text{MeV}$, and photon index $\Gamma = 2.1\pm 0.1\stat$. This spectrum is much harder than the typical spectrum of \source, and its normalization is much reduced. 
The change in flux and shape is clearly visible in Fig.~\ref{fig:spectra}(top).

In order to verify that the change in spectral shape coincided with the flux drop, two more power-law spectra have been derived for the time ranges MJD~59397-59411 and MJD~59415-59429 on either side of 2021 July 18 (MJD~59413).
The spectral indices are $2.57 \pm 0.09\stat$ and $2.1 \pm 0.1\stat$, respectively. These are compatible with the spectral shapes obtained for the longer periods confirming that the spectrum changed at the same time as the flux dropped.

\subsection{X-rays} \label{sec:swift}
%
\textit{Swift} \citep{Gehrels04} is a multi-frequency observatory for the X-ray and optical domain.
X-ray data in the energy range of 0.3-10\,keV collected with the X-ray Telescope \citep[XRT,][]{Burrows05} have been analyzed from 2021 and 2022, corresponding to the ObsIDs 00030797022-00030797027 and  00031173220-00030797029. 
They were taken in photon counting mode. 
The data analysis was performed using the HEASOFT software  (version 6.31), while for the recalibration the standard \verb|xrtpipeline| procedure was used.
\verb|xspec| \citep{Arnaud96} was employed for the spectral fitting.
All observations have been binned so that each bin contains at least 30 counts and each individual observation has been fitted with a single power-law model with a Galactic absorption value of $N_{H} = 7.13 \E{20}$\,cm$^{-2}$ \citep{HI4PI} set as a frozen parameter. 

The XRT light curve is shown in Fig.~\ref{fig:mwl_lightcurve}(c). The flux is consistent with being constant in 2021. The average flux in 2022 is reduced by less than a factor 2 compared to 2021, even though the flux varies mildly around the average (see Tab.~\ref{tab:xrtdata}). The average spectral shapes of 2021 and 2022 are very similar (see Fig.~\ref{fig:spectra}, middle and bottom, and Tab.~\ref{tab:xrtdata}).

\subsection{Optical/UV data} \label{sec:atom}
\subsubsection{Photometry}
%
%
Optical/UV photometry data have been collected with the Ultraviolet/Optical Telescope \citep[UVOT,][]{Roming05} onboard \textit{Swift} in six filters --- UVW2~(192.8\,nm), UVM2~(224.6\,nm), UVW1~(260.0\,nm), U~(346.5\,nm), B~(439.2\,nm), and V~(546.8\,nm) \citep{Poole08} --- as well as with the Automatic Telescope for Optical Monitoring \citep[ATOM, a 75\,cm aperture instrument located on the H.E.S.S. site,][]{hauser+04} with high cadence in BR filters.
%
For UVOT, magnitudes and corresponding fluxes have been calculated using \verb|uvotsource| including all photons from a circular region with radius 5''. In order to determine the background, a circular region with a radius of 10'' located near the source area has been selected. All data points are corrected for dust absorption using the reddening $E(B-V)$ = 0.0853\,mag  \citep{schlaflyfinkbeiner11} and the ratios of the extinction to reddening, $A_{\lambda} / E(B-V)$ from \cite{Giommi06}. 
%
%
The ATOM data were analysed using the fully automated ATOM Data Reduction and Analysis Software and their quality has been checked manually. 
The resulting flux was calculated via differential photometry using five custom-calibrated secondary standard stars in the same field of view. Extinction correction was done as for \textit{Swift}-UVOT.

The light curves in R and B filters are shown in Fig.~\ref{fig:mwl_lightcurve}(d). While variability is clearly visible in the 2021 data, the 2022 light curves show no significant variations. The fractional variability in the R- and B-band in 2022 is $3\%$ and $2\%$, respectively. The change in behavior seems to occur near-simultaneously with the flux drop in the HE \g-ray band, but the data is very sparse after July 2021, which is why a firm conclusion cannot be drawn. Interestingly, the R-B color also shows variability [see Fig.~\ref{fig:mwl_lightcurve}(e)]. In the high flux states in 2021, the R-band flux is higher than the B-band flux, while it is inverted for the low flux states, which is especially noticeable in 2022. In terms of B-R color, this change happens at B-R$\approx 0.6\,$mag.

For the spectra shown in Fig.~\ref{fig:spectra} and~\ref{fig:LEAverage}, fluxes in given filters have been averaged within the observation range of H.E.S.S., namely MJD 59311–59382 for 2021 and MJD 59672–59794 for 2022. While this includes some variability in 2021, it does not, for instance, include the peak in early July.
Nonetheless, the high variability in 2021 results in an average of the ATOM data that cannot be properly compared to the \textit{Swift}-UVOT averages, which were taken on at most six occasions and not necessarily parallel to the ATOM data. Therefore in Sec.~\ref{sec:res}, the R-band average from ATOM is treated as an upper limit for the 2021 data set, while the spetral fitting is done on the V, B, U, and UVM2 bands of \textit{Swift}-UVOT.

\subsubsection{Spectropolarimetry} \label{sec:salt}
%
Optical spectropolarimetric observations of \source\ were taken with the Southern African Large Telescope \citep[SALT,][]{2006_SALT_Buckley}, using the Robert Stobie Spectrograph \citep[RSS,][]{2003SPIE.4841.1463B, 2003SPIE.4841.1634K}. \source\ was observed eight times between 2021 April 06 and 2021 June 10, and eleven times between 2022 April 25 and 2022 July 31. All observations were performed using grating PG0900 at a grating angle of $12.875^{\circ}$ with a slit width of 1.25'' giving a resolving power of $R \approx 800 - 1200$. Observations were performed in {\sc linear} mode which takes four observations at 4 wave plate angles. A total exposure time of 1200\,s ($4\times300$\,s) was used for the first eight observations, and 1440\,s ($4\times360$\,s) for the remaining observations.
Data reduction was performed using a modified version of the pySALT/polSALT pipeline \citep{2010SPIE.7737E..25C}\footnote{\url{https://github.com/saltastro/polsalt}} allowing for the wavelength calibration to be performed with {\sc IRAF}\footnote{Version 2.16} \citep[see][]{cooper22}.  

The average degree of polarization was calculated for each observation in four different wavelength bands (see Fig.~\ref{fig:mwl_lightcurve}(f) and (g)), namely $\lambda = 3670 - 4060\,$\AA, $\lambda = 4100 - 4400\,$\AA, $\lambda = 4480 - 4780\,$\AA, and $\lambda = 4800 - 5100\,$\AA, chosen to avoid spectral features. During the 2021 observing period, the source exhibited variable levels of polarization, reaching a maximum of $\langle \Pi \rangle = 12.5 \pm 1.1\,\%$ on 2021 May 08 (taken between $\lambda = 4100 - 6200\,$\AA), and a minimum of $\langle \Pi \rangle = 2.2 \pm 0.5\,\%$ on 2021 April 20. During the 2021 semester, the polarization angle varied by $\sim\,174^{\circ}$ (reaching a maximum of $178.9 \pm 4.8^{\circ}$ on 2021 April 20, and a minimum of $4.7 \pm 2.7^{\circ}$ on 2021 April 09).

During 2022, the source exhibited little to no variation in the degree of polarization, consistently remaining below $2\,\%$. This is consistent with the level of polarization measured for a comparison star. Thus, the observed polarization can be attributed to interstellar effects, rather than any source-intrinsic polarization.

%
%
\section{Results} \label{sec:res}
\begin{figure*}
\centering
\includegraphics[width=0.96\textwidth]{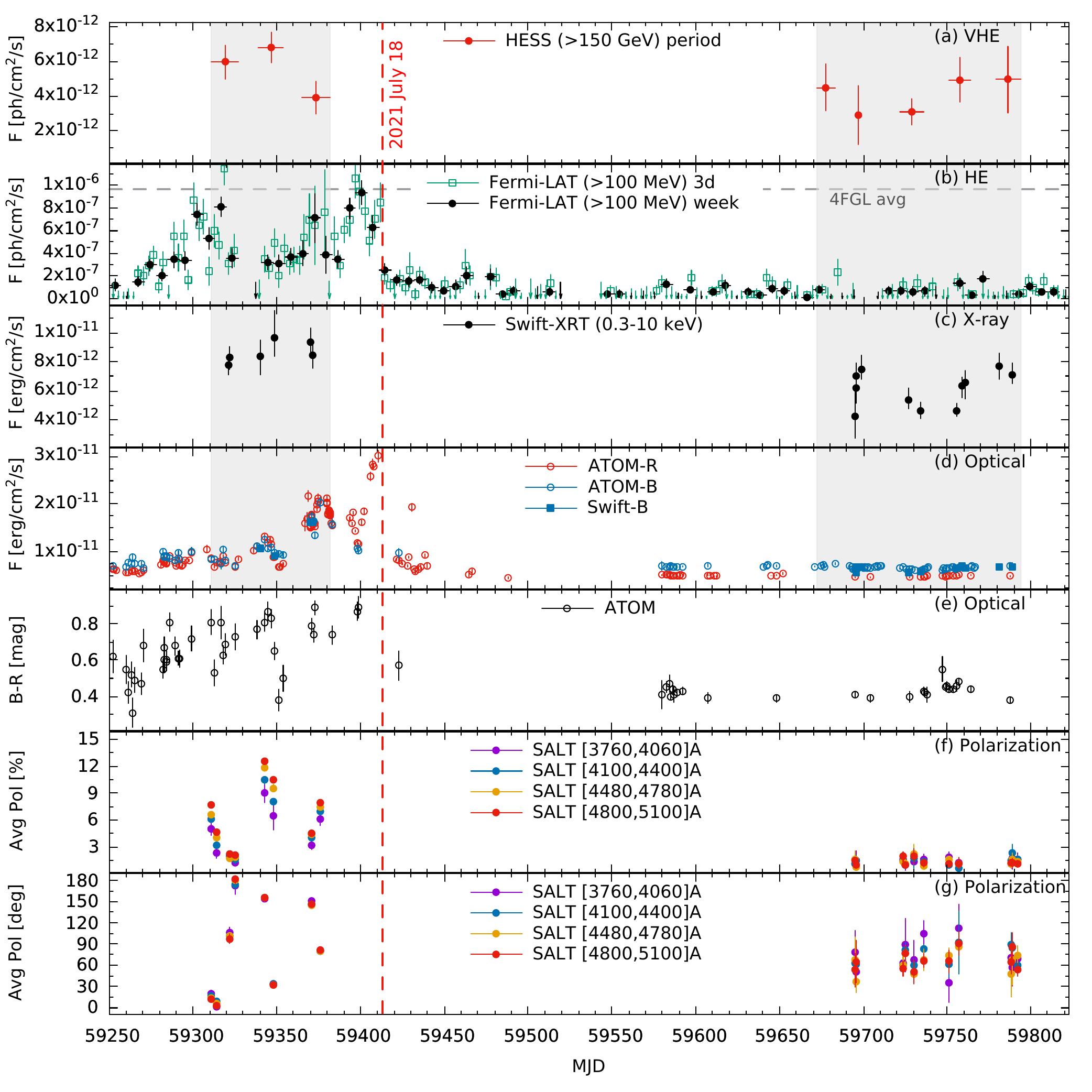}
\caption{MWL light curves of PKS~1510-089 in 2021 and 2022. 
\textbf{(a)} Period-wise VHE \g-ray light curve from \hess\ integrated above an energy threshold of $150\,$GeV showing statistical errors only. 
\textbf{(b)} HE \g-ray light curve from \fermi\ integrated above an energy threshold of $100\,$MeV in 3\,d (green open squares) and 7\,d (black filled circles) bins. Arrows display upper limits. The gray dashed line marks the average flux of the 4FGL-DR3 catalog. 
\textbf{(c)} X-ray light curve from \textit{Swift}-XRT integrated between $0.3$ and $10\,$keV for each observation. 
\textbf{(d)} Optical light curve from ATOM and \textit{Swift}-UVOT in BR filters for individual observations. 
\textbf{(e)} Optical B-R color from ATOM. 
\textbf{(f)} Average polarization degree and \textbf{(g)} average polarization angle from SALT in optical bands as indicated for each observation. 
The vertical red dashed line marks 2021 July 18 (MJD~59413). The gray shaded regions mark the time frames for which the average SEDs have been derived.
}
\label{fig:mwl_lightcurve}
\end{figure*}
\begin{figure}
\centering
\includegraphics[width=0.405\textwidth]{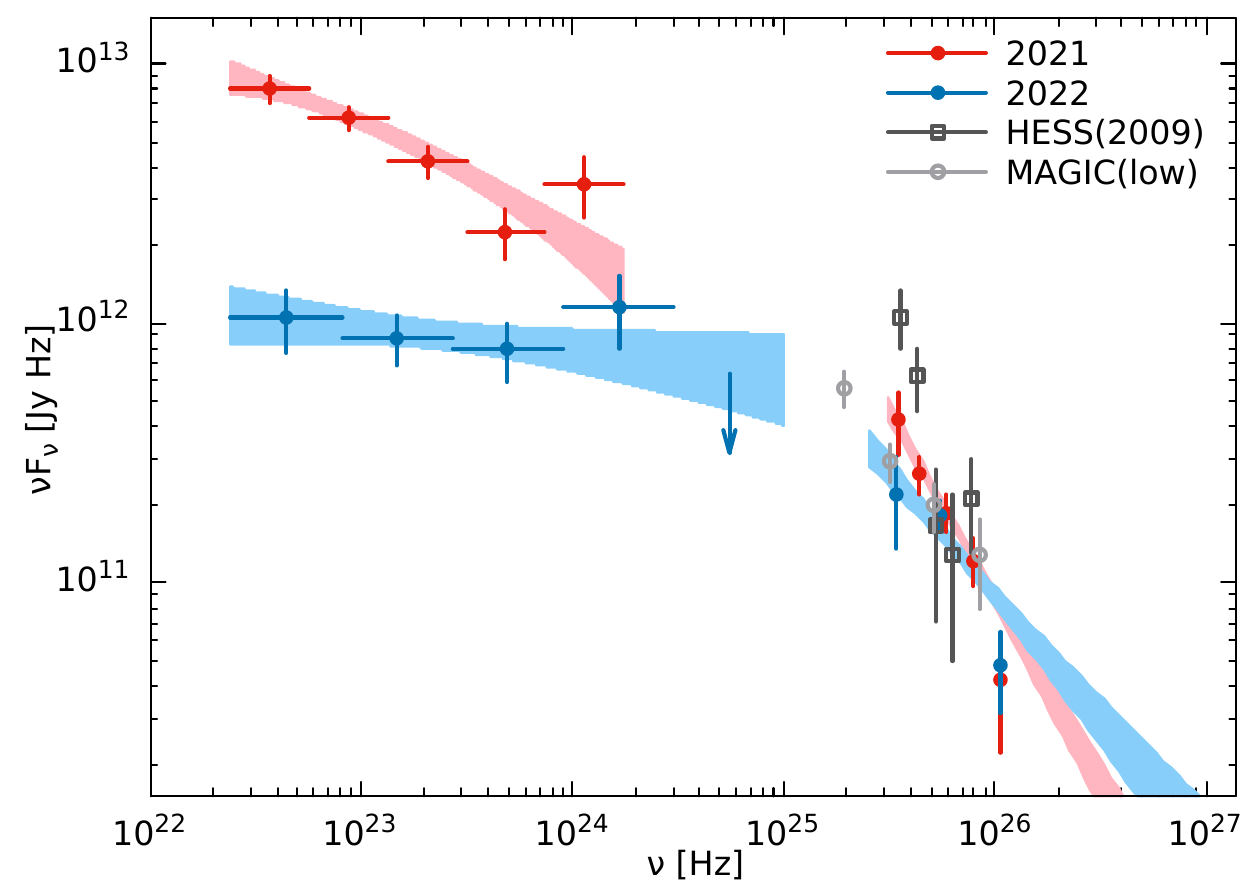}\\
\includegraphics[width=0.405\textwidth]{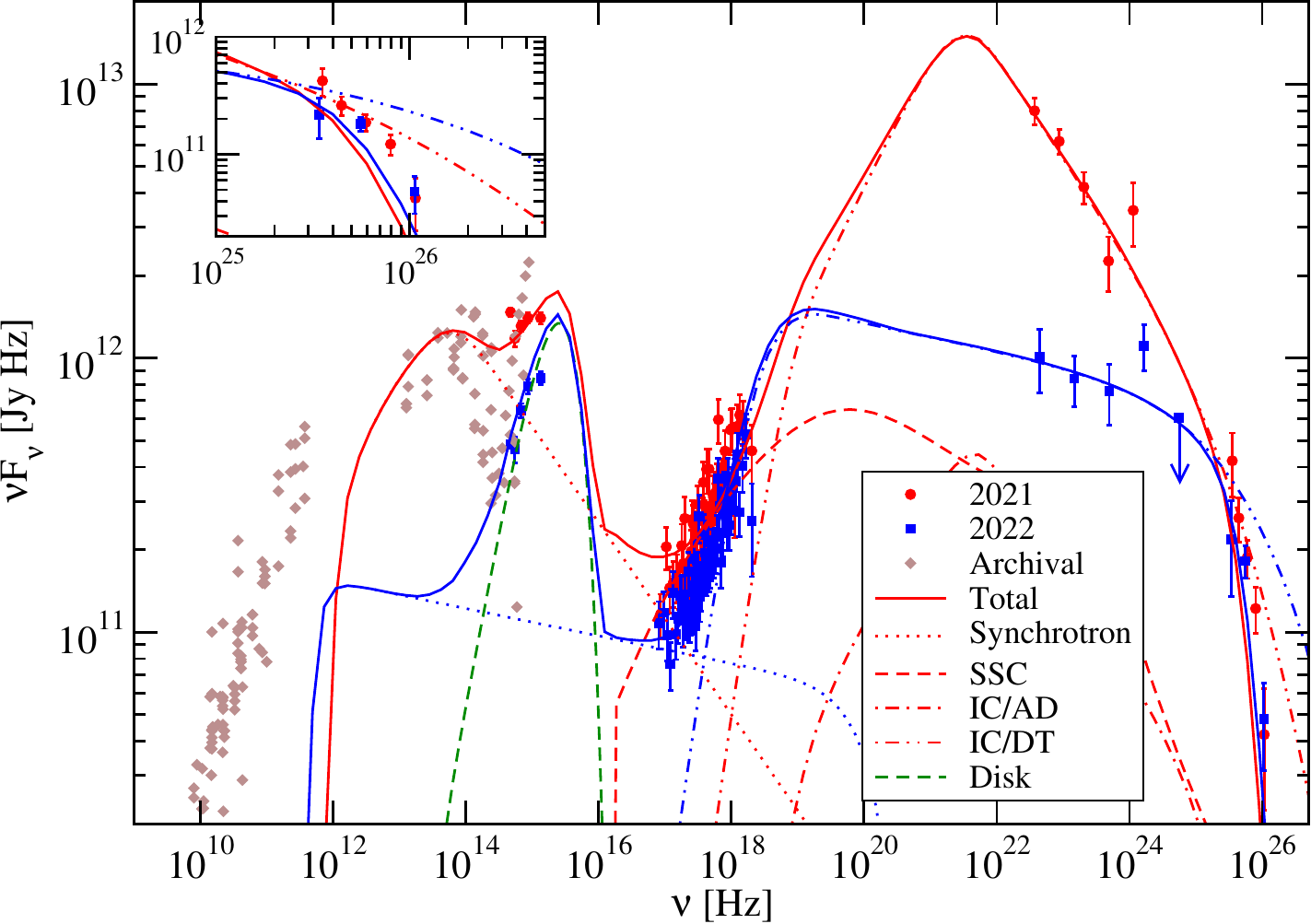}\\
\includegraphics[width=0.405\textwidth]{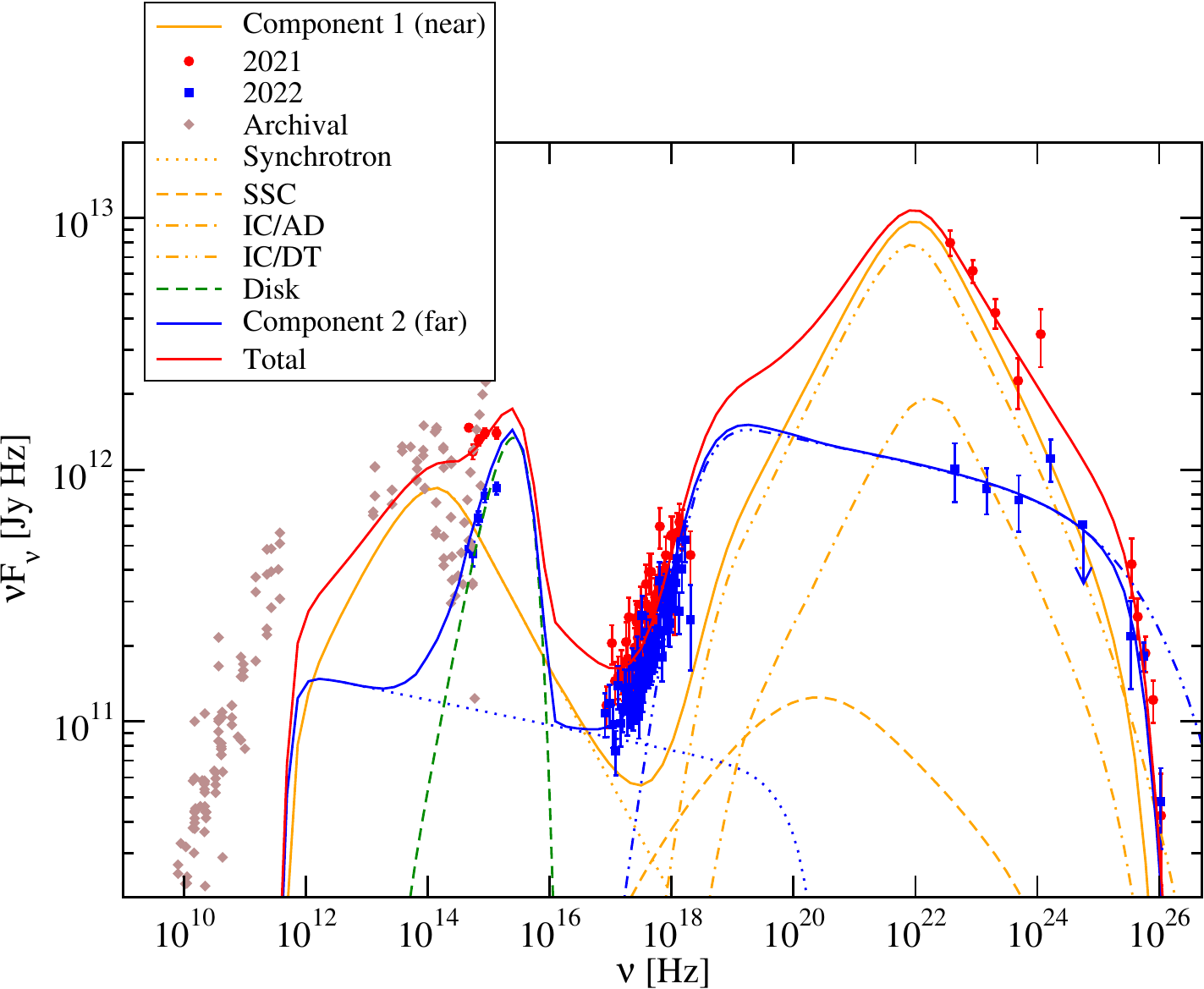}
\caption{\textit{Top:} Observed HE and VHE \g-ray spectrum for the 2021 (red) and 2022 (blue) data sets showing statistical errors only. The dark gray open squares are from \cite{hess13}, while the light gray open circles mark the spectrum from \cite{magic18}. 
\textit{Middle:} Leptonic single-zone model fits to the SED of PKS~1510-089 in 2021 (MJD 59311-59382, red) and 2022 (MJD 59672-59794, blue). 
The inset at the top left shows a zoom-in to the VHE $\gamma$-ray spectrum, illustrating the difficulty in finding a satisfactory model fit. 
\textit{Bottom:} Leptonic two-zone model fits to the SED of PKS~1510-089 in 2021 (MJD 59311-59382, red) and 2022 (MJD 59672-59794, blue).
In the middle and bottom panels, extinction corrections are applied to optical and X-ray fluxes but not to the \g-ray data, while the total model lines account for the EBL absorption. Gray points mark archival data taken from the NED (\url{http://ned.ipac.caltech.edu/}) considered as upper limits for the modeling.
}
\label{fig:spectra}
\end{figure}
The MWL light curves and spectra of \source\ are shown in Figs.~\ref{fig:mwl_lightcurve} and \ref{fig:spectra}, respectively. They show the aforementioned change in the source: most notably the HE \g-ray flux drop and spectral change, as well as the optical flux and polarization drop. These took place at a seemingly singular event around 2021 July 18 (MJD~59413). Interestingly, the VHE \g-ray and X-ray fluxes and spectra barely changed (within a factor 2), and the VHE \g-ray spectrum is a smooth continuation of the HE \g-ray spectrum in both years. The drop in optical polarization, along with the R-B color change, suggests that the optical-UV spectrum is strongly dominated by the AD and the BLR. 
In order to explore this further, a joint fit of the low-frequency SED and the optical spectropolarimetry is produced first to constrain the relative contributions of the jet synchrotron emission, the accretion-disk, and emission lines from the BLR as well as the jet emission-region parameters related to synchrotron emission (radiating relativistic electron distribution and magnetic field --- see Sec.~\ref{sec:opt}). The resulting parameters are then used in a second step to model the entire broadband SED, including X-rays and $\gamma$-rays, constraining additional parameters pertaining to the target photon fields for inverse-Compton scattering (Sec.~\ref{sec:mod}).

\subsection{Modeling the Optical-UV photometry and spectropolarimetry}\label{sec:opt}
Generally, the degree of polarization of the optical-UV jet synchrotron emission is diluted by the non-polarized, thermal contributions of the AD and the BLR.
%
The model of \cite{Schutte_2022} (see also App.~\ref{sec:app1} for further details) derives the synchrotron state of a blazar assuming a single emission zone 
containing an electron distribution

\begin{align}
 N_e (\gamma) = n_0 
  \left\lbrace \begin{array}{r@{}l@{\qquad}l}
    & {}(\frac{\gamma}{\gamma_b})^{-p_1} \cdot e^{-\gamma_b/ \gamma_c} \textrm{ for}\ \gamma_{\rm min} \le \gamma \le \gamma_{\rm b}, \\[\jot]
    & {}(\frac{\gamma}{\gamma_b})^{-p_2} \cdot e^{-\gamma/\gamma_c} \textrm{ for}\ \gamma_{\rm b} \le \gamma_{\rm max},
  \end{array}\right. 
 \label{eq:edistr}
\end{align}
with electron spectral indices $p_1$ and $p_2$ where, in the slow-cooling regime, one expects $p_2 = p_1 + 1$. The characteristic Lorentz factors are in the range $[\gamma_{\rm min}, \gamma_{\rm max}]$ with a break of a broken power-law spectrum at $\gamma_b$  and an exponential cut-off at $\gamma_{\rm c}$.
Synchrotron self-absorption effects are also considered.
The model implements a geometrically thin, optically thick AD \citep{shakurasunyaev73} around a non-rotating supermassive black hole of mass $M_{BH} = 6 \times 10^8 \rm \, M_{\sun}$, which is within the range of previously obtained mass estimates, $5.71^{+0.62}_{-0.58} \times 10^7 \rm \; M_{\odot}$ and $7 \times 10^8 \rm \; M_{\odot}$, by \cite{Rakshit2020} and \cite{Ghisellinietal2010}, respectively. 
For an AD accretion rate $\dot{M_d}$, the efficiency of converting potential energy into AD radiation is assumed to be $\epsilon = L_d/(\dot{M_d} c^2) = 1/12$ \citep{Ghisellinietal2010}. The different states from 2021 to 2022 can be modeled with an unchanging AD. 

The synchrotron polarization was calculated following \citet[Eq.~(6.38)]{1979rpa..book.....R}. The degree of polarization depends on the geometry of the magnetic field in the jet. This is characterized by the scaling factor $F_B$ between 0 and 1, with $1$ representing perfectly ordered magnetic fields, whereas values less than 1 represent more tangled magnetic fields. The total degree of polarization is calculated as the sum of the synchrotron polarization and the unpolarized AD and BLR emissions.

The emission lines can be modeled as Gaussian functions and the corresponding fluxes can be calculated relative to each other according to \cite{Phillips1978}. 
Their model did not include the H$\alpha$, C IV and Ly$\alpha$ lines. However, these were considered by \cite{Malkanetal1986} and \cite{Isleretal2015} 
alongside the Mg II, H$\gamma$, H$\beta$ and H$\alpha$ emission lines. 
The CIV, Mg II, H$\gamma$ and H$\alpha$ emission lines are also included here, while emission lines are excluded if they are outside of the frequency regime with good spectropolarimetric or photometric data. 

\begin{figure*}
\includegraphics[width=0.96\textwidth]{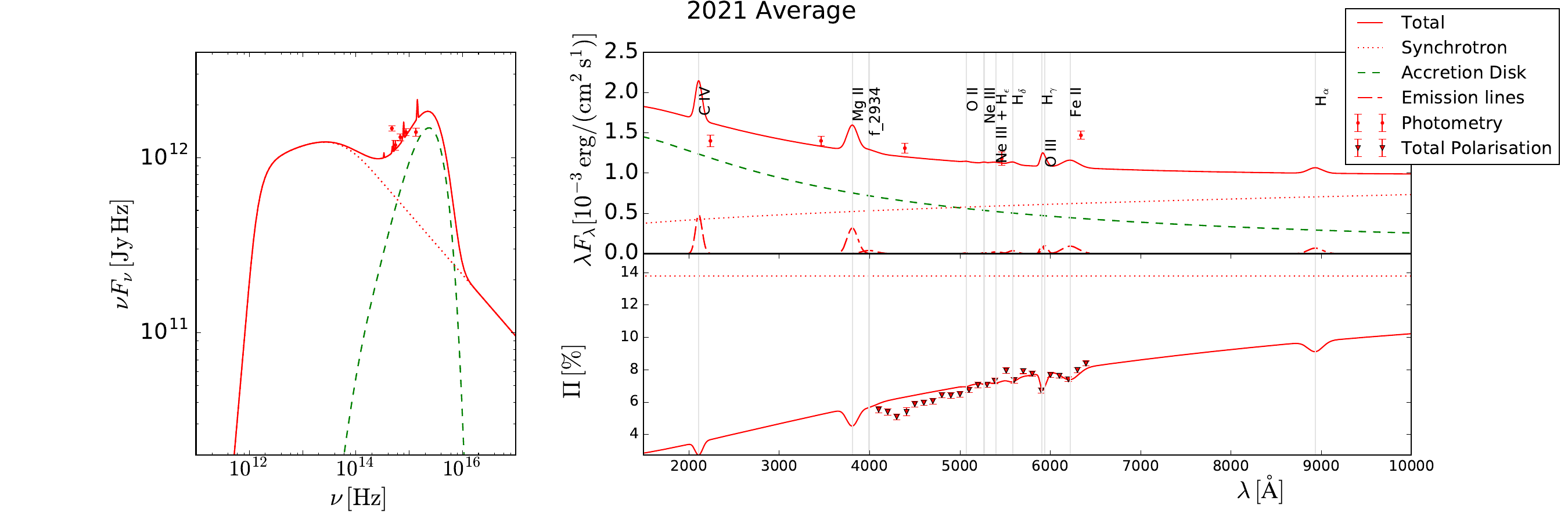}\\
\includegraphics[width=0.96\textwidth]{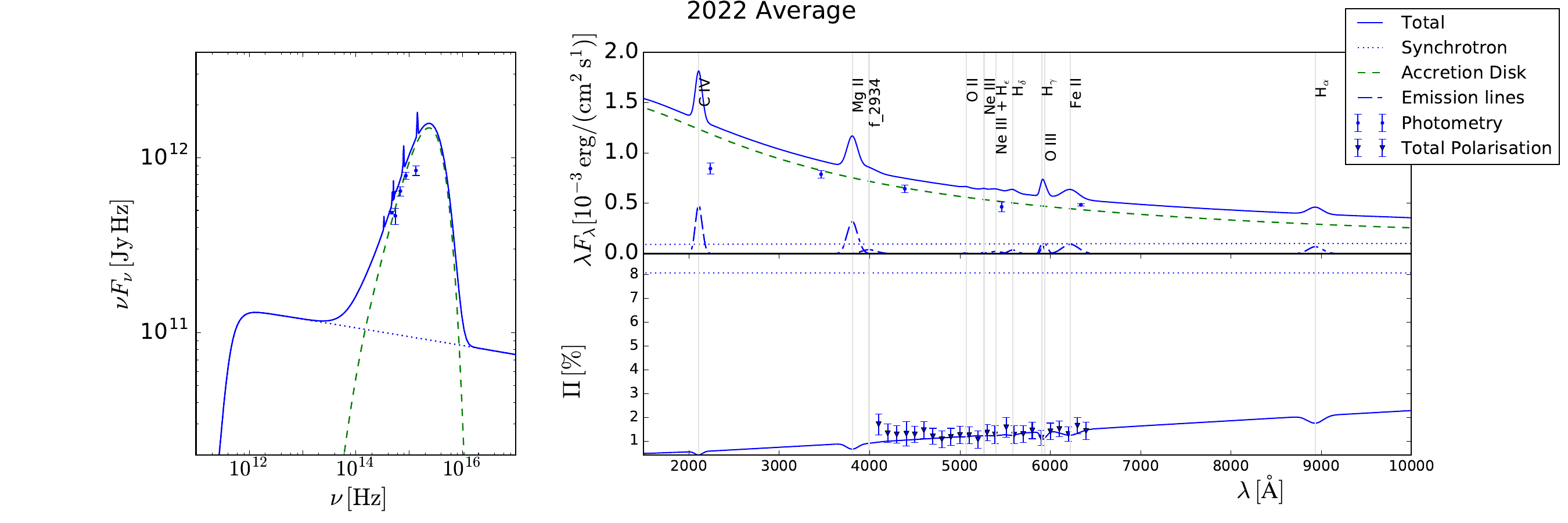}
\caption{Application of the \cite{Schutte_2022} model to the data averaged over 2021 (red) and 2022 (blue) with data points as in Fig.~\ref{fig:spectra} (middle and bottom). The first SED bump is shown in the left panels and the components in the optical-UV regime are shown in the upper-right panels with corresponding polarization in the lower-right panels. The plotted radiation components are: Synchrotron (dotted), AD flux (green dashed), emission lines (double-dash-dotted) and the total of the components (solid). Individual emission line fluxes are not plotted in the left panels; they are, however, plotted in the upper-right panels.
}
\label{fig:LEAverage}
\end{figure*}

The data averaged over 2021 and 2022 are modeled and shown in Fig.~\ref{fig:LEAverage}. In 2021, there are contributions by synchrotron, AD, and BLR radiation, while the data in 2022 requires dominating AD and BLR flux. The upper-right panel, showing the 2022 fit, suggests that the photometry data can be well fitted with only the AD and line components without the synchrotron contribution. 
Thus, the fit to the 2022 data marks a strict upper limit to the synchrotron flux contribution, 
in line with the above statement that the source-intrinsic polarization is consistent with zero in \source. 
The parameters obtained with the model fits are given in Tab.~\ref{tab:OptUVpars}. 
%
For reference, the model application to all individual observations in 2021 is shown in appendix~\ref{sec:app1}.

The simultaneous modeling of the flux and polarization shows that the jet's synchrotron emission must have dropped considerably between 2021 and 2022, leaving behind the AD and the BLR as the almost sole flux contributors in the optical/UV regime. This underlines the unprecedented change that took place in \source.

\begin{table*}
    \centering
    \caption{Parameters obtained by fitting the optical/UV flux and polarization data averaged over 2021 and 2022, respectively, with the code of \cite{Schutte_2022}. Constant parameters are: Bulk Lorentz factor $\Gamma = 20$, $\theta_{obs} = 2.9\,$deg, magnetic field $B = 2\,$G, emission region radius $R = 3 \times 10^{15}\,$cm, $\gamma_{min} = 1$, accretion disk luminosity $L_{d} = 1.8 \times 10^{46}\,$erg\,s$^{-1}$ and $M_{BH} = 6 \times 10^8 \rm \, M_{\sun}$. $n_0$ is the electron distribution normalization. The $\chi^2_{pol}/ndf$ is the goodness of fit to the spectropolarimetry data, with degrees of freedom $ndf = 13$.}\label{tab:OptUVpars}
    \begin{tabular}{l ccccccc}
    \hline
        Date & $n_0$ & $\gamma_b$ & $\gamma_c$ &  $p_1$ &  $p_2$ & $F_B$ & $\chi^2_{pol}/ndf $ \\  \hline
        2021 Average &  $1 \times 10^{47}$ & 569 & $5.0 \times 10^{6}$ & 2.7 & 3.7 & 0.18 & 0.06 \\ 
        2022 Average & $7 \times 10^{49}$ & 30 & $5.0 \times 10^{6}$ & 2.1 & 3.1 & 0.1 & 0.10 \\ \hline 
    \end{tabular}
\end{table*}

\subsection{Broadband SED modeling} \label{sec:mod}
%
In this section, first a fit of the broadband (IR -- VHE $\gamma$-ray) SEDs of \source\ of 2021 and 2022 is attempted with a simple one-zone, steady-state leptonic model. For this purpose, the leptonic code of \cite{Boettcher+13} is employed. See that paper for a detailed description of the model, which includes IC scattering of the co-spatially produced synchrotron emission (SSC) and external Compton scattering of the AD emission (IC/AD), modeled with the parameters derived in Sec.~\ref{sec:opt}, and of the DT, modeled as an isotropic (in the AGN rest frame) blackbody photon field (IC/DT). The most relevant model parameters are thus: The injection luminosity of non-thermal electrons, $L_{\rm inj}$, the low- and high-energy cut-offs of the injected electron spectrum, $\gamma_{\rm min}$ and $\gamma_{\rm max}$, the electron injection spectral index $p_1$, the size of the emission region, $R$, the co-moving magnetic field $B$, the bulk Lorentz factor $\Gamma$, the viewing angle $\theta_{\rm obs}$ (in the observer's frame), the distance of the emission region from the black hole, $z_0$, and the energy density and equivalent temperature of the external blackbody radiation field, $u_{\rm ext}$ and $T_{\rm ext}$. The code evaluates self-consistently an equilibrium electron distribution, based on the balance between injection/acceleration, radiative cooling, and escape, evaluates the kinetic jet power $L_e$ corresponding to the final electron population in the emission region and the Poynting flux power $L_B$, and calculates the ratio $L_B/L_e = u_B/u_e$ which provides information on the magnetization of the jet plasma. The absorption through the extragalactic background light is evaluated with the model of \cite{finke+10}.
Given the large number of parameters, a fit by eye is conducted, as a proper $\chi^2$ minimization procedure is not feasible, and it would likely be degenerate in any case, since many of the model parameters are very poorly constrained. 


Fig.~\ref{fig:spectra}(middle) shows representative attempts of single-zone leptonic fits to the 2021 (red) and 2022 (blue) SEDs. The adopted model parameters are listed in Tab.~\ref{tab:parameters} 
and are chosen in such a way that the resulting radiating electron distribution is identical to the one resulting from the low-frequency SED and spectropolarimetry fit in Sec.~\ref{sec:opt}. 
The distance of the emission region in 2021 is very poorly constrained, as a small contribution of IC/AD emission slightly improves the fit, but is not strictly required. An almost identical fit can be achieved with a much larger distance from the black hole, assuming that the DT radiation field has the same energy density at that distance. The soft HE \g-ray spectrum, implying a very soft electron spectrum, combined with Klein-Nishina effects at the highest energies, makes it very difficult to find a satisfactory fit to the \hess\ spectral points in this single-zone scenario. 

For the 2022 low state, the HE \g-ray and non-thermal optical flux may be suppressed by using a smaller injection luminosity / acceleration efficiency and a significantly harder injection spectrum. In order to suppress any potential contribution of IC/AD, a distance $z_0 \gg 0.1\,$pc from the black hole is required. The parameters adopted for the 2022 single-zone fit shown in Fig.~\ref{fig:spectra} (middle) have been chosen to keep as many parameters as possible unchanged between 2021 and 2022. However, if the dominant emission region in 2022 is indeed much further down the jet than in 2021, keeping the magnetic field and emission-region radius constant may not be plausible. A fit with a decreased magnetic field (such as $B \propto z_0^{-1}$, as expected for a dominantly toroidal magnetic field) and larger emission region (such as $R \propto z_0$ for a conical jet) leads to an almost identical fit to the X-ray through VHE $\gamma$-ray flux, but strongly suppresses the synchrotron emission in the radio through X-ray regime. 

Due to the difficulty of finding a satisfactory fit to the VHE spectrum in 2021, now the possibility of a two-zone model is explored, which is shown in Fig.~\ref{fig:spectra}(bottom). As the X-ray and VHE $\gamma$-ray spectra appear to have remained almost unchanged between 2021 and 2022, it seems natural to postulate a steady emission region responsible for the non-thermal emission in 2022, which may have been active also in 2021, with the additional emission region, closer to the central engine, that was only active in 2021. Therefore, the parameters of the far zone equal to the 2022 SED fit described above are kept, while a near zone is added with parameters listed in the last column of Tab.~\ref{tab:parameters}. This produces a satisfactory fit to the entire SED in 2021 (including the \hess\ points) with physical conditions close to equipartition (bottom row in Tab.~\ref{tab:parameters}) in both emission regions. It should be noted that the B-field in the far-zone (2022) is poorly constrained and could easily be chosen to achieve exact equipartition. 

Absorption of \g\ rays in circum-nuclear radiation fields (accretion-disk, BLR) has not been accounted for in the model fits. It has been shown by \cite{Reimer07} for strong-lined AGN in general and by \cite{BE16} specifically for \source\ that VHE $\gamma$-rays are expected to be strongly attenuated if the emission region were located at sub-pc distances from the central engine. The fact that the VHE spectrum of \source\ does not show any signs of such internal $\gamma\gamma$ absorption \citep[see also][]{hess+21} provides further support for the far-zone interpretation. 
This goes in line with the choice not to add an EC/BLR radiation component to the far-zone model. Such a component could plausibly be present in the near-zone / 2021 model. However, the IC/DT spectrum provides a satisfactory fit to the Fermi-LAT spectrum in 2021, and an IC/BLR component would not significantly contribute to the VHE spectrum due to Klein-Nishina effects. Therefore, it is preferred not to include additional parameters to the model.

\begin{table}
\centering
\caption{\label{tab:parameters} Model parameters for the SED fits shown in Fig. \ref{fig:spectra} (middle and bottom). 
}
\begin{tabular}{cccc}
\hline
Parameter & 2021    & 2022   & 2021     \cr
[units]   & single-zone  & single-zone & two-zone \cr
\hline
$L_e$ [erg\,s$^{-1}$]       & $6.2 \times 10^{44}$ & $2.1 \times 10^{44} $ & $2.3 \times 10^{44}$ \cr
$\gamma_{\rm min}$           & 600                  & 30                    & $1.0 \times 10^3$    \cr
$\gamma_{\rm max}$           & $5.0 \times 10^6$    & $1.0 \times 10^6$     & $1.0 \times 10^6$    \cr 
$p_1$                        & 2.7                  & 2.1                   & 2.9                  \cr
$R$ [cm]                     & $3.0 \times 10^{15}$ & $1.0 \times 10^{16}$  & $5.0 \times 10^{15}$ \cr 
$B$ [G]                      & 2.0                  & 2.0                   & 2.2                  \cr 
$z_0$ [pc]                   & 0.1                  & 10                    & 0.06                 \cr 
$\Gamma$                     & 20                   & 20                    & 20                   \cr 
$\theta_{\rm obs}$ [deg]     & 2.9                  & 2.9                   & 2.9                  \cr 
$u_{\rm ext}$ [erg\,cm$^{-3}$]    & $1.5 \times 10^{-3}$ & $1.5 \times 10^{-3}$  & $1.5 \times 10^{-3}$ \cr 
$T_{\rm ext}$ [K]            & 100                  & 100                   & 100                  \cr 
$L_B$ [erg\,s$^{-1}$]        & $6.5 \times 10^{43}$ & $6.0 \times 10^{44}$  & $1.5 \times 10^{44}$ \cr
$L_B/L_e$                    & 0.11 & 2.8                   & 0.66                 \cr
\hline
\end{tabular}
\end{table}

%
%
\section{Discussion \& conclusions} \label{sec:con}
The relativistic jet of \source\ underwent a sudden and significant change around 2021 July 18. The HE \g-ray and optical fluxes observed with \fermi\ and ATOM, respectively, dropped to persistent low states, while the optical spectropolarimetry data obtained with SALT suggests a drop to a level compatible with no polarization in the source. The optical spectrum is thus fully explained by the AD and the BLR. Meanwhile, the VHE \g-ray and X-ray fluxes observed with H.E.S.S. and \textit{Swift}-XRT, respectively, remained steady within a factor 2.

This favors the two-zone interpretation, where separate emission regions were active before 2021 July 18 contributing to various degrees in all energy bands. Around this date, the primary zone close to the black hole that was responsible for most of the optical synchrotron and HE \g-ray emission, vanished leaving behind the secondary zone that has contributed strongly to the VHE \g-ray and X-ray domains. 
The secondary zone has been modeled as IC/DT at a few parsec from the black hole. In comparison to the two-zone interpretation in \cite{nalewajko+12}, a softer electron distribution and a slightly higher $\gamma_{\rm min}$ is required for the secondary zone described here owing to the different characteristics in the HE \g-ray domain. 
The new \g-ray state can also be reproduced with an IC/CMB model in the kpc-scale jet similar to \cite{Meyer+19}\footnote{There is a notable spectral difference in the HE \g-ray low-state spectrum in \cite{Meyer+19} compared to the one presented here, which suggests that the primary zone was active in the date set of \cite{Meyer+19}.} with the caveat that it cannot account for the X-ray spectrum measured with \textit{Swift}. 
The comparison of the current VHE \g-ray spectrum with the discovery spectrum [see Fig.~\ref{fig:spectra}(top)] suggests that the secondary zone was already present in the old data, but that the VHE spectrum was also influenced by the primary zone allowing for the reproduction of that data with a single-zone model \citep[e.g.,][]{barnacka+14}.
However, the two-zone explanation as outlined here would also explain the varying correlation patterns observed between the HE and VHE \g-ray bands \citep{zacharias+19}. 


The disappearance of the primary emission zone suggests two probable explanations. Either the inner jet has weakened considerably and is no longer capable of producing significant amounts of radiation, or the inner jet has swung away from the line-of-sight reducing the amount of Doppler beaming. 
Both scenarios may also explain the sudden termination of the flare that was ongoing in the HE and optical bands. In order to uncover the details of this event, elaborate modeling is required, which is beyond the scope of this paper.
In either case, the disturbance should be transported through the jet and may eventually reach the parsec-scale jet. On these scales, the changes become observable in VLBI radio maps by a reduced total flux, by an outward motion of the core (if the jet weakens and becomes incapable of producing radio flux at the current core position) or a gradual swing of the jet structure. 
Publicly available radio data\footnote{Such as from Metsähovi, \url{https://www.metsahovi.fi/AGN/data/}, and ATCA, \url{https://www.narrabri.atnf.csiro.au/calibrators/}, among others.} show a flare occuring around the time of the disappearance of the primary emission region. 
This suggests a connection, but a detailed analysis is left to future work.
Eventually, both scenarios could lead to a vanishing of the secondary emission zone, 
which could be uncovered in continuous MWL monitoring observations.

%
%
\begin{acknowledgments}
We thank the referee for a constructive report that helped to improve the manuscript.\\
The support of the Namibian authorities and of the University of
Namibia in facilitating the construction and operation of H.E.S.S.
is gratefully acknowledged, as is the support by the German
Ministry for Education and Research (BMBF), the Max Planck Society,
the German Research Foundation (DFG), the Helmholtz Association,
the Alexander von Humboldt Foundation, the French Ministry of
Higher Education, Research and Innovation, the Centre National de
la Recherche Scientifique (CNRS/IN2P3 and CNRS/INSU), the
Commissariat à l’énergie atomique et aux énergies alternatives
(CEA), the U.K. Science and Technology Facilities Council (STFC),
the Irish Research Council (IRC) and the Science Foundation Ireland
(SFI), the Knut and Alice Wallenberg Foundation, the Polish
Ministry of Education and Science, agreement no. 2021/WK/06, the
South African Department of Science and Technology and National
Research Foundation, the University of Namibia, the National
Commission on Research, Science \& Technology of Namibia (NCRST),
the Austrian Federal Ministry of Education, Science and Research
and the Austrian Science Fund (FWF), the Australian Research
Council (ARC), the Japan Society for the Promotion of Science, the
University of Amsterdam and the Science Committee of Armenia grant
21AG-1C085. We appreciate the excellent work of the technical
support staff in Berlin, Zeuthen, Heidelberg, Palaiseau, Paris,
Saclay, Tübingen and in Namibia in the construction and operation
of the equipment. This work benefited from services provided by the
H.E.S.S. Virtual Organisation, supported by the national resource
providers of the EGI Federation.\\
Some of the observations reported in this paper were obtained with the Southern African Large Telescope (SALT) under program 2021-2-LSP-001 (PI: D.A.H.~Buckley).\\
This research has made use of the NASA/IPAC Extragalactic Database (NED), which is funded by the National Aeronautics and Space Administration and operated by the California Institute of Technology.
\end{acknowledgments}
%
\vspace{5mm}


%
%
%
\bibliographystyle{aasjournal}
\bibliography{references}

\begin{thebibliography}{}
\expandafter\ifx\csname natexlab\endcsname\relax\def\natexlab#1{#1}\fi
\providecommand{\url}[1]{\href{#1}{#1}}
\providecommand{\dodoi}[1]{doi:~\href{http://doi.org/#1}{\nolinkurl{#1}}}
\providecommand{\doeprint}[1]{\href{http://ascl.net/#1}{\nolinkurl{http://ascl.net/#1}}}
\providecommand{\doarXiv}[1]{\href{https://arxiv.org/abs/#1}{\nolinkurl{https://arxiv.org/abs/#1}}}

\bibitem[{{Abdollahi} {et~al.}(2020){Abdollahi}, {Acero}, {Ackermann},
  {Ajello}, {Atwood}, {Axelsson}, {Baldini}, {Ballet}, {Barbiellini},
  {Bastieri}, {Becerra Gonzalez}, {Bellazzini}, {Berretta}, {Bissaldi},
  {Blandford}, {Bloom}, {Bonino}, {Bottacini}, {Brandt}, {Bregeon}, {Bruel},
  {Buehler}, {Burnett}, {Buson}, {Cameron}, {Caputo}, {Caraveo}, {Casandjian},
  {Castro}, {Cavazzuti}, {Charles}, {Chaty}, {Chen}, {Cheung}, {Chiaro},
  {Ciprini}, {Cohen-Tanugi}, {Cominsky}, {Coronado-Bl{\'a}zquez}, {Costantin},
  {Cuoco}, {Cutini}, {D'Ammando}, {DeKlotz}, {de la Torre Luque}, {de Palma},
  {Desai}, {Digel}, {Di Lalla}, {Di Mauro}, {Di Venere}, {Dom{\'\i}nguez},
  {Dumora}, {Fana Dirirsa}, {Fegan}, {Ferrara}, {Franckowiak}, {Fukazawa},
  {Funk}, {Fusco}, {Gargano}, {Gasparrini}, {Giglietto}, {Giommi}, {Giordano},
  {Giroletti}, {Glanzman}, {Green}, {Grenier}, {Griffin}, {Grondin}, {Grove},
  {Guiriec}, {Harding}, {Hayashi}, {Hays}, {Hewitt}, {Horan},
  {J{\'o}hannesson}, {Johnson}, {Kamae}, {Kerr}, {Kocevski}, {Kovac'evic'},
  {Kuss}, {Landriu}, {Larsson}, {Latronico}, {Lemoine-Goumard}, {Li},
  {Liodakis}, {Longo}, {Loparco}, {Lott}, {Lovellette}, {Lubrano}, {Madejski},
  {Maldera}, {Malyshev}, {Manfreda}, {Marchesini}, {Marcotulli},
  {Mart{\'\i}-Devesa}, {Martin}, {Massaro}, {Mazziotta}, {McEnery}, {Mereu},
  {Meyer}, {Michelson}, {Mirabal}, {Mizuno}, {Monzani}, {Morselli},
  {Moskalenko}, {Negro}, {Nuss}, {Ojha}, {Omodei}, {Orienti}, {Orlando},
  {Ormes}, {Palatiello}, {Paliya}, {Paneque}, {Pei}, {Pe{\~n}a-Herazo},
  {Perkins}, {Persic}, {Pesce-Rollins}, {Petrosian}, {Petrov}, {Piron}, {Poon},
  {Porter}, {Principe}, {Rain{\`o}}, {Rando}, {Razzano}, {Razzaque}, {Reimer},
  {Reimer}, {Remy}, {Reposeur}, {Romani}, {Saz Parkinson}, {Schinzel},
  {Serini}, {Sgr{\`o}}, {Siskind}, {Smith}, {Spandre}, {Spinelli}, {Strong},
  {Suson}, {Tajima}, {Takahashi}, {Tak}, {Thayer}, {Thompson}, {Tibaldo},
  {Torres}, {Torresi}, {Valverde}, {Van Klaveren}, {van Zyl}, {Wood},
  {Yassine}, \& {Zaharijas}}]{abdollahi20}
{Abdollahi}, S., {Acero}, F., {Ackermann}, M., {et~al.} 2020, \apjs, 247, 33,
  \dodoi{10.3847/1538-4365/ab6bcb}

\bibitem[{{Acero} {et~al.}(2016){Acero}, {Ackermann}, {Ajello}, {Albert},
  {Baldini}, {Ballet}, {Barbiellini}, {Bastieri}, {Bellazzini}, {Bissaldi},
  {Bloom}, {Bonino}, {Bottacini}, {Brandt}, {Bregeon}, {Bruel}, {Buehler},
  {Buson}, {Caliandro}, {Cameron}, {Caragiulo}, {Caraveo}, {Casandjian},
  {Cavazzuti}, {Cecchi}, {Charles}, {Chekhtman}, {Chiang}, {Chiaro}, {Ciprini},
  {Claus}, {Cohen-Tanugi}, {Conrad}, {Cuoco}, {Cutini}, {D'Ammando}, {de
  Angelis}, {de Palma}, {Desiante}, {Digel}, {Di Venere}, {Drell}, {Favuzzi},
  {Fegan}, {Ferrara}, {Focke}, {Franckowiak}, {Funk}, {Fusco}, {Gargano},
  {Gasparrini}, {Giglietto}, {Giordano}, {Giroletti}, {Glanzman}, {Godfrey},
  {Grenier}, {Guiriec}, {Hadasch}, {Harding}, {Hayashi}, {Hays}, {Hewitt},
  {Hill}, {Horan}, {Hou}, {Jogler}, {J{\'o}hannesson}, {Kamae}, {Kuss},
  {Landriu}, {Larsson}, {Latronico}, {Li}, {Li}, {Longo}, {Loparco},
  {Lovellette}, {Lubrano}, {Maldera}, {Malyshev}, {Manfreda}, {Martin},
  {Mayer}, {Mazziotta}, {McEnery}, {Michelson}, {Mirabal}, {Mizuno}, {Monzani},
  {Morselli}, {Nuss}, {Ohsugi}, {Omodei}, {Orienti}, {Orlando}, {Ormes},
  {Paneque}, {Pesce-Rollins}, {Piron}, {Pivato}, {Rain{\`o}}, {Rando},
  {Razzano}, {Razzaque}, {Reimer}, {Reimer}, {Remy}, {Renault},
  {S{\'a}nchez-Conde}, {Schaal}, {Schulz}, {Sgr{\`o}}, {Siskind}, {Spada},
  {Spandre}, {Spinelli}, {Strong}, {Suson}, {Tajima}, {Takahashi}, {Thayer},
  {Thompson}, {Tibaldo}, {Tinivella}, {Torres}, {Tosti}, {Troja}, {Vianello},
  {Werner}, {Wood}, {Wood}, {Zaharijas}, \& {Zimmer}}]{acero16}
{Acero}, F., {Ackermann}, M., {Ajello}, M., {et~al.} 2016, \apjs, 223, 26,
  \dodoi{10.3847/0067-0049/223/2/26}

\bibitem[{{Aharonian} {et~al.}(2006){Aharonian}, {Akhperjanian}, {Bazer-Bachi},
  {Beilicke}, {Benbow}, {Berge}, {Bernl{\"o}hr}, {Boisson}, {Bolz}, {Borrel},
  {Braun}, {Breitling}, {Brown}, {B{\"u}hler}, {B{\"u}sching}, {Carrigan},
  {Chadwick}, {Chounet}, {Cornils}, {Costamante}, {Degrange}, {Dickinson},
  {Djannati-Ata{\"\i}}, {O'C. Drury}, {Dubus}, {Egberts}, {Emmanoulopoulos},
  {Espigat}, {Feinstein}, {Ferrero}, {Fiasson}, {Fontaine}, {Funk}, {Funk},
  {Gallant}, {Giebels}, {Glicenstein}, {Goret}, {Hadjichristidis}, {Hauser},
  {Hauser}, {Heinzelmann}, {Henri}, {Hermann}, {Hinton}, {Hofmann}, {Holleran},
  {Horns}, {Jacholkowska}, {de Jager}, {Kh{\'e}lifi}, {Komin}, {Konopelko},
  {Kosack}, {Latham}, {Le Gallou}, {Lemi{\`e}re}, {Lemoine-Goumard}, {Lohse},
  {Martin}, {Martineau-Huynh}, {Marcowith}, {Masterson}, {McComb}, {de
  Naurois}, {Nedbal}, {Nolan}, {Noutsos}, {Orford}, {Osborne}, {Ouchrif},
  {Panter}, {Pelletier}, {Pita}, {P{\"u}hlhofer}, {Punch}, {Raubenheimer},
  {Raue}, {Rayner}, {Reimer}, {Reimer}, {Ripken}, {Rob}, {Rolland}, {Rowell},
  {Sahakian}, {Saug{\'e}}, {Schlenker}, {Schlickeiser}, {Schwanke}, {Sol},
  {Spangler}, {Spanier}, {Steenkamp}, {Stegmann}, {Superina}, {Tavernet},
  {Terrier}, {Th{\'e}oret}, {Tluczykont}, {van Eldik}, {Vasileiadis}, {Venter},
  {Vincent}, {V{\"o}lk}, {Wagner}, \& {Ward}}]{hess06}
{Aharonian}, F., {Akhperjanian}, A.~G., {Bazer-Bachi}, A.~R., {et~al.} 2006,
  \aap, 457, 899, \dodoi{10.1051/0004-6361:20065351}

\bibitem[{{Ajello} {et~al.}(2020){Ajello}, {Angioni}, {Axelsson}, {Ballet},
  {Barbiellini}, {Bastieri}, {Becerra Gonzalez}, {Bellazzini}, {Bissaldi},
  {Bloom}, {Bonino}, {Bottacini}, {Bruel}, {Buson}, {Cafardo}, {Cameron},
  {Cavazzuti}, {Chen}, {Cheung}, {Ciprini}, {Costantin}, {Cutini}, {D'Ammando},
  {de la Torre Luque}, {de Menezes}, {de Palma}, {Desai}, {Di Lalla}, {Di
  Venere}, {Dom{\'\i}nguez}, {Dirirsa}, {Ferrara}, {Finke}, {Franckowiak},
  {Fukazawa}, {Funk}, {Fusco}, {Gargano}, {Garrappa}, {Gasparrini},
  {Giglietto}, {Giordano}, {Giroletti}, {Green}, {Grenier}, {Guiriec},
  {Harita}, {Hays}, {Horan}, {Itoh}, {J{\'o}hannesson}, {Kovac'evic'},
  {Krauss}, {Kreter}, {Kuss}, {Larsson}, {Leto}, {Li}, {Liodakis}, {Longo},
  {Loparco}, {Lott}, {Lovellette}, {Lubrano}, {Madejski}, {Maldera},
  {Manfreda}, {Mart{\'\i}-Devesa}, {Massaro}, {Mazziotta}, {Mereu}, {Meyer},
  {Migliori}, {Mirabal}, {Mizuno}, {Monzani}, {Morselli}, {Moskalenko},
  {Negro}, {Nemmen}, {Nuss}, {Ojha}, {Ojha}, {Omodei}, {Orienti}, {Orlando},
  {Ormes}, {Paliya}, {Pei}, {Pe{\~n}a-Herazo}, {Persic}, {Pesce-Rollins},
  {Petrov}, {Piron}, {Poon}, {Principe}, {Rain{\`o}}, {Rando}, {Rani},
  {Razzano}, {Razzaque}, {Reimer}, {Reimer}, {Schinzel}, {Serini}, {Sgr{\`o}},
  {Siskind}, {Spandre}, {Spinelli}, {Suson}, {Tachibana}, {Thompson}, {Torres},
  {Torresi}, {Troja}, {Valverde}, {van Zyl}, \& {Yassine}}]{ajello20}
{Ajello}, M., {Angioni}, R., {Axelsson}, M., {et~al.} 2020, \apj, 892, 105,
  \dodoi{10.3847/1538-4357/ab791e}

\bibitem[{{Arnaud}(1996)}]{Arnaud96}
{Arnaud}, K.~A. 1996, in Astronomical Society of the Pacific Conference Series,
  Vol. 101, Astronomical Data Analysis Software and Systems V, ed. G.~H.
  {Jacoby} \& J.~{Barnes}, 17

\bibitem[{{Atwood} {et~al.}(2009){Atwood}, {Abdo}, {Ackermann}, {Althouse},
  {Anderson}, {Axelsson}, {Baldini}, {Ballet}, {Band}, {Barbiellini},
  {Bartelt}, {Bastieri}, {Baughman}, {Bechtol}, {B{\'e}d{\'e}r{\`e}de},
  {Bellardi}, {Bellazzini}, {Berenji}, {Bignami}, {Bisello}, {Bissaldi},
  {Blandford}, {Bloom}, {Bogart}, {Bonamente}, {Bonnell}, {Borgland},
  {Bouvier}, {Bregeon}, {Brez}, {Brigida}, {Bruel}, {Burnett}, {Busetto},
  {Caliandro}, {Cameron}, {Caraveo}, {Carius}, {Carlson}, {Casandjian},
  {Cavazzuti}, {Ceccanti}, {Cecchi}, {Charles}, {Chekhtman}, {Cheung},
  {Chiang}, {Chipaux}, {Cillis}, {Ciprini}, {Claus}, {Cohen-Tanugi},
  {Condamoor}, {Conrad}, {Corbet}, {Corucci}, {Costamante}, {Cutini}, {Davis},
  {Decotigny}, {DeKlotz}, {Dermer}, {de Angelis}, {Digel}, {do Couto e Silva},
  {Drell}, {Dubois}, {Dumora}, {Edmonds}, {Fabiani}, {Farnier}, {Favuzzi},
  {Flath}, {Fleury}, {Focke}, {Funk}, {Fusco}, {Gargano}, {Gasparrini},
  {Gehrels}, {Gentit}, {Germani}, {Giebels}, {Giglietto}, {Giommi}, {Giordano},
  {Glanzman}, {Godfrey}, {Grenier}, {Grondin}, {Grove}, {Guillemot}, {Guiriec},
  {Haller}, {Harding}, {Hart}, {Hays}, {Healey}, {Hirayama}, {Hjalmarsdotter},
  {Horn}, {Hughes}, {J{\'o}hannesson}, {Johansson}, {Johnson}, {Johnson},
  {Johnson}, {Johnson}, {Kamae}, {Katagiri}, {Kataoka}, {Kavelaars}, {Kawai},
  {Kelly}, {Kerr}, {Klamra}, {Kn{\"o}dlseder}, {Kocian}, {Komin}, {Kuehn},
  {Kuss}, {Landriu}, {Latronico}, {Lee}, {Lee}, {Lemoine-Goumard}, {Lionetto},
  {Longo}, {Loparco}, {Lott}, {Lovellette}, {Lubrano}, {Madejski}, {Makeev},
  {Marangelli}, {Massai}, {Mazziotta}, {McEnery}, {Menon}, {Meurer},
  {Michelson}, {Minuti}, {Mirizzi}, {Mitthumsiri}, {Mizuno}, {Moiseev},
  {Monte}, {Monzani}, {Moretti}, {Morselli}, {Moskalenko}, {Murgia},
  {Nakamori}, {Nishino}, {Nolan}, {Norris}, {Nuss}, {Ohno}, {Ohsugi}, {Omodei},
  {Orlando}, {Ormes}, {Paccagnella}, {Paneque}, {Panetta}, {Parent}, {Pearce},
  {Pepe}, {Perazzo}, {Pesce-Rollins}, {Picozza}, {Pieri}, {Pinchera}, {Piron},
  {Porter}, {Poupard}, {Rain{\`o}}, {Rando}, {Rapposelli}, {Razzano}, {Reimer},
  {Reimer}, {Reposeur}, {Reyes}, {Ritz}, {Rochester}, {Rodriguez}, {Romani},
  {Roth}, {Russell}, {Ryde}, {Sabatini}, {Sadrozinski}, {Sanchez}, {Sander},
  {Sapozhnikov}, {Parkinson}, {Scargle}, {Schalk}, {Scolieri}, {Sgr{\`o}},
  {Share}, {Shaw}, {Shimokawabe}, {Shrader}, {Sierpowska-Bartosik}, {Siskind},
  {Smith}, {Smith}, {Spandre}, {Spinelli}, {Starck}, {Stephens}, {Strickman},
  {Strong}, {Suson}, {Tajima}, {Takahashi}, {Takahashi}, {Tanaka}, {Tenze},
  {Tether}, {Thayer}, {Thayer}, {Thompson}, {Tibaldo}, {Tibolla}, {Torres},
  {Tosti}, {Tramacere}, {Turri}, {Usher}, {Vilchez}, {Vitale}, {Wang},
  {Watters}, {Winer}, {Wood}, {Ylinen}, \& {Ziegler}}]{atwood09}
{Atwood}, W.~B., {Abdo}, A.~A., {Ackermann}, M., {et~al.} 2009, \apj, 697,
  1071, \dodoi{10.1088/0004-637X/697/2/1071}

\bibitem[{{Barnacka} {et~al.}(2014){Barnacka}, {Moderski}, {Behera}, {Brun}, \&
  {Wagner}}]{barnacka+14}
{Barnacka}, A., {Moderski}, R., {Behera}, B., {Brun}, P., \& {Wagner}, S. 2014,
  \aap, 567, A113, \dodoi{10.1051/0004-6361/201322205}

\bibitem[{{Boettcher} {et~al.}(2012){Boettcher}, {Harris}, \&
  {Krawczynski}}]{boettcherHarrisKrawczynski12}
{Boettcher}, M., {Harris}, D.~E., \& {Krawczynski}, H. 2012, {Relativistic Jets
  from Active Galactic Nuclei} (Wiley-VCH)

\bibitem[{{B{\"o}ttcher}(2019)}]{boettcher19}
{B{\"o}ttcher}, M. 2019, Galaxies, 7, 20, \dodoi{10.3390/galaxies7010020}

\bibitem[{{B{\"o}ttcher} \& {Els}(2016)}]{BE16}
{B{\"o}ttcher}, M., \& {Els}, P. 2016, \apj, 821, 102,
  \dodoi{10.3847/0004-637X/821/2/102}

\bibitem[{{B{\"o}ttcher} {et~al.}(2013){B{\"o}ttcher}, {Reimer}, {Sweeney}, \&
  {Prakash}}]{Boettcher+13}
{B{\"o}ttcher}, M., {Reimer}, A., {Sweeney}, K., \& {Prakash}, A. 2013, \apj,
  768, 54, \dodoi{10.1088/0004-637X/768/1/54}

\bibitem[{{Brown}(2013)}]{brown13}
{Brown}, A.~M. 2013, \mnras, 431, 824, \dodoi{10.1093/mnras/stt218}

\bibitem[{Buckley {et~al.}(2006)Buckley, Swart, \& Meiring}]{2006_SALT_Buckley}
Buckley, D., Swart, G., \& Meiring, J. 2006, Proceedings of SPIE - The
  International Society for Optical Engineering, 6267,
  \dodoi{10.1117/12.673750}

\bibitem[{{Burbidge} \& {Kinman}(1966)}]{burbidgekinman66}
{Burbidge}, E.~M., \& {Kinman}, T.~D. 1966, \apj, 145, 654,
  \dodoi{10.1086/148808}

\bibitem[{{Burgh} {et~al.}(2003){Burgh}, {Nordsieck}, {Kobulnicky}, {Williams},
  {O'Donoghue}, {Smith}, \& {Percival}}]{2003SPIE.4841.1463B}
{Burgh}, E.~B., {Nordsieck}, K.~H., {Kobulnicky}, H.~A., {et~al.} 2003, in
  Society of Photo-Optical Instrumentation Engineers (SPIE) Conference Series,
  Vol. 4841, Instrument Design and Performance for Optical/Infrared
  Ground-based Telescopes, ed. M.~{Iye} \& A.~F.~M. {Moorwood}, 1463--1471,
  \dodoi{10.1117/12.460312}

\bibitem[{{Burrows} {et~al.}(2005){Burrows}, {Hill}, {Nousek}, {Kennea},
  {Wells}, {Osborne}, {Abbey}, {Beardmore}, {Mukerjee}, {Short}, {Chincarini},
  {Campana}, {Citterio}, {Moretti}, {Pagani}, {Tagliaferri}, {Giommi},
  {Capalbi}, {Tamburelli}, {Angelini}, {Cusumano}, {Br{\"a}uninger}, {Burkert},
  \& {Hartner}}]{Burrows05}
{Burrows}, D.~N., {Hill}, J.~E., {Nousek}, J.~A., {et~al.} 2005, \ssr, 120,
  165, \dodoi{10.1007/s11214-005-5097-2}

\bibitem[{{Cerruti}(2020)}]{cerruti20}
{Cerruti}, M. 2020, Galaxies, 8, 72, \dodoi{10.3390/galaxies8040072}

\bibitem[{{Cooper} {et~al.}(2022){Cooper}, {van Soelen}, \&
  {Britto}}]{cooper22}
{Cooper}, J., {van Soelen}, B., \& {Britto}, R. 2022, in High Energy
  Astrophysics in Southern Africa 2021, 56, \dodoi{10.22323/1.401.0056}

\bibitem[{{Crawford} {et~al.}(2010){Crawford}, {Still}, {Schellart}, {Balona},
  {Buckley}, {Dugmore}, {Gulbis}, {Kniazev}, {Kotze}, {Loaring}, {Nordsieck},
  {Pickering}, {Potter}, {Romero Colmenero}, {Vaisanen}, {Williams}, \&
  {Zietsman}}]{2010SPIE.7737E..25C}
{Crawford}, S.~M., {Still}, M., {Schellart}, P., {et~al.} 2010, in Society of
  Photo-Optical Instrumentation Engineers (SPIE) Conference Series, Vol. 7737,
  Observatory Operations: Strategies, Processes, and Systems III, ed. D.~R.
  {Silva}, A.~B. {Peck}, \& B.~T. {Soifer}, 773725, \dodoi{10.1117/12.857000}

\bibitem[{{de Naurois} \& {Rolland}(2009)}]{denauroisrolland09}
{de Naurois}, M., \& {Rolland}, L. 2009, Astroparticle Physics, 32, 231,
  \dodoi{10.1016/j.astropartphys.2009.09.001}

\bibitem[{{Finke} {et~al.}(2010){Finke}, {Razzaque}, \& {Dermer}}]{finke+10}
{Finke}, J.~D., {Razzaque}, S., \& {Dermer}, C.~D. 2010, \apj, 712, 238,
  \dodoi{10.1088/0004-637X/712/1/238}

\bibitem[{{Francis} {et~al.}(1991){Francis}, {Hewett}, {Foltz}, {Chaffee},
  {Weymann}, \& {Morris}}]{Francisetal1991}
{Francis}, P.~J., {Hewett}, P.~C., {Foltz}, C.~B., {et~al.} 1991, ApJ, 373,
  465, \dodoi{10.1086/170066}

\bibitem[{{Gehrels} {et~al.}(2004){Gehrels}, {Chincarini}, {Giommi}, {Mason},
  {Nousek}, {Wells}, {White}, {Barthelmy}, {Burrows}, {Cominsky}, {Hurley},
  {Marshall}, {M{\'e}sz{\'a}ros}, {Roming}, {Angelini}, {Barbier}, {Belloni},
  {Campana}, {Caraveo}, {Chester}, {Citterio}, {Cline}, {Cropper}, {Cummings},
  {Dean}, {Feigelson}, {Fenimore}, {Frail}, {Fruchter}, {Garmire}, {Gendreau},
  {Ghisellini}, {Greiner}, {Hill}, {Hunsberger}, {Krimm}, {Kulkarni}, {Kumar},
  {Lebrun}, {Lloyd-Ronning}, {Markwardt}, {Mattson}, {Mushotzky}, {Norris},
  {Osborne}, {Paczynski}, {Palmer}, {Park}, {Parsons}, {Paul}, {Rees},
  {Reynolds}, {Rhoads}, {Sasseen}, {Schaefer}, {Short}, {Smale}, {Smith},
  {Stella}, {Tagliaferri}, {Takahashi}, {Tashiro}, {Townsley}, {Tueller},
  {Turner}, {Vietri}, {Voges}, {Ward}, {Willingale}, {Zerbi}, \&
  {Zhang}}]{Gehrels04}
{Gehrels}, N., {Chincarini}, G., {Giommi}, P., {et~al.} 2004, \apj, 611, 1005,
  \dodoi{10.1086/422091}

\bibitem[{Ghisellini {et~al.}(2010)Ghisellini, Tavecchio, Foschini, Ghirlanda,
  Maraschi, \& Celotti}]{Ghisellinietal2010}
Ghisellini, G., Tavecchio, F., Foschini, L., {et~al.} 2010, Monthly Notices of
  the Royal Astronomical Society, 402, 497,
  \dodoi{10.1111/j.1365-2966.2009.15898.x}

\bibitem[{{Giommi} {et~al.}(2006){Giommi}, {Blustin}, {Capalbi},
  {Colafrancesco}, {Cucchiara}, {Fuhrmann}, {Krimm}, {Marchili}, {Massaro},
  {Perri}, {Tagliaferri}, {Tosti}, {Tramacere}, {Burrows}, {Chincarini},
  {Falcone}, {Gehrels}, {Kennea}, \& {Sambruna}}]{Giommi06}
{Giommi}, P., {Blustin}, A.~J., {Capalbi}, M., {et~al.} 2006, \aap, 456, 911,
  \dodoi{10.1051/0004-6361:20064874}

\bibitem[{{H.~E.~S.~S. Collaboration} {et~al.}(2013){H.~E.~S.~S.
  Collaboration}, {Abramowski}, {Acero}, {Aharonian}, {Akhperjanian}, {Anton},
  {Balenderan}, {Balzer}, {Barnacka}, {Becherini}, {Becker Tjus}, {Behera},
  {Bernl{\"o}hr}, {Birsin}, {Biteau}, {Bochow}, {Boisson}, {Bolmont}, {Bordas},
  {Brucker}, {Brun}, {Brun}, {Bulik}, {Carrigan}, {Casanova}, {Cerruti},
  {Chadwick}, {Chaves}, {Cheesebrough}, {Colafrancesco}, {Cologna}, {Conrad},
  {Couturier}, {Dalton}, {Daniel}, {Davids}, {Degrange}, {Deil}, {deWilt},
  {Dickinson}, {Djannati-Ata{\"\i}}, {Domainko}, {O'C. Drury}, {Dubus},
  {Dutson}, {Dyks}, {Dyrda}, {Egberts}, {Eger}, {Espigat}, {Fallon}, {Farnier},
  {Fegan}, {Feinstein}, {Fernandes}, {Fernandez}, {Fiasson}, {Fontaine},
  {F{\"o}rster}, {F{\"u}{\ss}ling}, {Gajdus}, {Gallant}, {Garrigoux}, {Gast},
  {Giebels}, {Glicenstein}, {Gl{\"u}ck}, {G{\"o}ring}, {Grondin},
  {Grudzi{\'n}ska}, {H{\"a}ffner}, {Hague}, {Hahn}, {Hampf}, {Harris},
  {Hauser}, {Heinz}, {Heinzelmann}, {Henri}, {Hermann}, {Hillert}, {Hinton},
  {Hofmann}, {Hofverberg}, {Holler}, {Horns}, {Jacholkowska}, {Jahn},
  {Jamrozy}, {Jung}, {Kastendieck}, {Katarzy{\'n}ski}, {Katz}, {Kaufmann},
  {Kh{\'e}lifi}, {Klepser}, {Klochkov}, {Klu{\'z}niak}, {Kneiske}, {Kolitzus},
  {Komin}, {Kosack}, {Kossakowski}, {Krayzel}, {Kr{\"u}ger}, {Laffon},
  {Lamanna}, {Lefaucheur}, {Lemoine-Goumard}, {Lenain}, {Lennarz}, {Lohse},
  {Lopatin}, {Lu}, {Marandon}, {Marcowith}, {Masbou}, {Maurin}, {Maxted},
  {Mayer}, {McComb}, {Medina}, {M{\'e}hault}, {Menzler}, {Moderski}, {Mohamed},
  {Moulin}, {Naumann}, {Naumann-Godo}, {de Naurois}, {Nedbal}, {Nguyen},
  {Niemiec}, {Nolan}, {Ohm}, {de O{\~n}a Wilhelmi}, {Opitz}, {Ostrowski},
  {Oya}, {Panter}, {Parsons}, {Paz Arribas}, {Pekeur}, {Pelletier}, {Perez},
  {Petrucci}, {Peyaud}, {Pita}, {P{\"u}hlhofer}, {Punch}, {Quirrenbach},
  {Raab}, {Raue}, {Reimer}, {Reimer}, {Renaud}, {de los Reyes}, {Rieger},
  {Ripken}, {Rob}, {Rosier-Lees}, {Rowell}, {Rudak}, {Rulten}, {Sahakian},
  {Sanchez}, {Santangelo}, {Schlickeiser}, {Schulz}, {Schwanke}, {Schwarzburg},
  {Schwemmer}, {Sheidaei}, {Skilton}, {Sol}, {Spengler}, {Stawarz},
  {Steenkamp}, {Stegmann}, {Stinzing}, {Stycz}, {Sushch}, {Szostek},
  {Tavernet}, {Terrier}, {Tluczykont}, {Trichard}, {Valerius}, {van Eldik},
  {Vasileiadis}, {Venter}, {Viana}, {Vincent}, {V{\"o}lk}, {Volpe}, {Vorobiov},
  {Vorster}, {Wagner}, {Ward}, {White}, {Wierzcholska}, {Wouters}, {Zacharias},
  {Zajczyk}, {Zdziarski}, {Zech}, \& {Zechlin}}]{hess13}
{H.~E.~S.~S. Collaboration}, {Abramowski}, A., {Acero}, F., {et~al.} 2013,
  \aap, 554, A107, \dodoi{10.1051/0004-6361/201321135}

\bibitem[{{Hauser} {et~al.}(2004){Hauser}, {M{\"o}llenhoff}, {P{\"u}hlhofer},
  {Wagner}, {Hagen}, \& {Knoll}}]{hauser+04}
{Hauser}, M., {M{\"o}llenhoff}, C., {P{\"u}hlhofer}, G., {et~al.} 2004,
  Astronomische Nachrichten, 325, 659, \dodoi{10.1002/asna.200410305}

\bibitem[{{H.E.S.S. Collaboration} {et~al.}(2021){H.E.S.S. Collaboration},
  {Abdalla}, {Adam}, {Aharonian}, {Ait Benkhali}, {Ang{\"u}ner}, {Arcaro},
  {Armand}, {Armstrong}, {Ashkar}, {Backes}, {Baghmanyan}, {Barbosa Martins},
  {Barnacka}, {Barnard}, {Becherini}, {Berge}, {Bernl{\"o}hr}, {Bi},
  {B{\"o}ttcher}, {Boisson}, {Bolmont}, {Bonnefoy}, {de Bony de Lavergne},
  {Bregeon}, {Breuhaus}, {Brun}, {Brun}, {Bryan}, {B{\"u}chele}, {Bulik},
  {Bylund}, {Caroff}, {Carosi}, {Casanova}, {Chand}, {Chandra}, {Chen},
  {Cotter}, {Cury{\l}o}, {Damascene Mbarubucyeye}, {Davids}, {Davies}, {Deil},
  {Devin}, {Dewilt}, {Dirson}, {Djannati-Ata{\"\i}}, {Dmytriiev}, {Donath},
  {Doroshenko}, {Dyks}, {Egberts}, {Eichhorn}, {Einecke}, {Emery}, {Ernenwein},
  {Feijen}, {Fegan}, {Fiasson}, {Fichet de Clairfontaine}, {Filipovic},
  {Fontaine}, {Funk}, {F{\"u}{\ss}ling}, {Gabici}, {Gallant}, {Giavitto},
  {Giunti}, {Glawion}, {Glicenstein}, {Gottschall}, {Grondin}, {Hahn}, {Haupt},
  {Hermann}, {Hinton}, {Hofmann}, {Hoischen}, {Holch}, {Holler}, {H{\"o}rbe},
  {Horns}, {Huber}, {Jamrozy}, {Jankowsky}, {Jankowsky}, {Jardin-Blicq},
  {Joshi}, {Jung-Richardt}, {Kastendieck}, {Katarzy{\'n}ski}, {Katz},
  {Khangulyan}, {Kh{\'e}lifi}, {Klepser}, {Klu{\'z}niak}, {Komin}, {Konno},
  {Kosack}, {Kostunin}, {Kreter}, {Lamanna}, {Lemi{\`e}re}, {Lemoine-Goumard},
  {Lenain}, {Levy}, {Lohse}, {Lypova}, {Mackey}, {Majumdar}, {Malyshev},
  {Malyshev}, {Marandon}, {Marchegiani}, {Marcowith}, {Mares},
  {Mart{\'\i}-Devesa}, {Marx}, {Maurin}, {Meintjes}, {Meyer}, {Mitchell},
  {Moderski}, {Mohamed}, {Mohrmann}, {Montanari}, {Moore}, {Morris}, {Moulin},
  {Muller}, {Murach}, {Nakashima}, {Nayerhoda}, {de Naurois}, {Ndiyavala},
  {Niederwanger}, {Niemiec}, {Oakes}, {O'Brien}, {Odaka}, {Ohm},
  {Olivera-Nieto}, {de Ona Wilhelmi}, {Ostrowski}, {Panter}, {Panny},
  {Parsons}, {Peron}, {Peyaud}, {Piel}, {Pita}, {Poireau}, {Priyana Noel},
  {Prokhorov}, {Prokoph}, {P{\"u}hlhofer}, {Punch}, {Quirrenbach}, {Raab},
  {Rauth}, {Reichherzer}, {Reimer}, {Reimer}, {Remy}, {Renaud}, {Rieger},
  {Rinchiuso}, {Romoli}, {Rowell}, {Rudak}, {Ruiz-Velasco}, {Sahakian},
  {Sailer}, {Sanchez}, {Santangelo}, {Sasaki}, {Scalici}, {Sch{\"u}ssler},
  {Schutte}, {Schwanke}, {Schwemmer}, {Seglar-Arroyo}, {Senniappan},
  {Seyffert}, {Shafi}, {Shiningayamwe}, {Simoni}, {Sinha}, {Sol}, {Specovius},
  {Spencer}, {Spir-Jacob}, {Stawarz}, {Sun}, {Steenkamp}, {Stegmann},
  {Steinmassl}, {Steppa}, {Takahashi}, {Tavernier}, {Taylor}, {Terrier},
  {Tiziani}, {Tluczykont}, {Tomankova}, {Trichard}, {Tsirou}, {Tuffs},
  {Uchiyama}, {van der Walt}, {van Eldik}, {van Rensburg}, {van Soelen},
  {Vasileiadis}, {Veh}, {Venter}, {Vincent}, {Vink}, {V{\"o}lk}, {Vuillaume},
  {Wadiasingh}, {Wagner}, {Watson}, {Werner}, {White}, {Wierzcholska}, {Wong},
  {Yusafzai}, {Zacharias}, {Zanin}, {Zargaryan}, {Zdziarski}, {Zech}, {Zhu},
  {Zorn}, {Zouari}, {{\.Z}ywucka}, {MAGIC Collaboration}, {Acciari}, {Ansoldi},
  {Antonelli}, {Arbet Engels}, {Asano}, {Baack}, {Babi{\'c}}, {Baquero},
  {Barres de Almeida}, {Barrio}, {Becerra Gonz{\'a}lez}, {Bednarek},
  {Bellizzi}, {Bernardini}, {Berti}, {Besenrieder}, {Bhattacharyya},
  {Bigongiari}, {Biland}, {Blanch}, {Bonnoli}, {Bo{\v{s}}njak}, {Busetto},
  {Carosi}, {Ceribella}, {Cerruti}, {Chai}, {Chilingarian}, {Cikota}, {Colak},
  {Colin}, {Colombo}, {Contreras}, {Cortina}, {Covino}, {D'Amico}, {D'Elia},
  {da Vela}, {Dazzi}, {de Angelis}, {de Lotto}, {Delfino}, {Delgado},
  {Depaoli}, {di Pierro}, {di Venere}, {Do Souto Espi{\~n}eira}, {Dominis
  Prester}, {Donini}, {Dorner}, {Doro}, {Elsaesser}, {Fallah Ramazani},
  {Fattorini}, {Ferrara}, {Foffano}, {Fonseca}, {Font}, {Fruck}, {Fukami},
  {Garc{\'\i}a L{\'o}pez}, {Garczarczyk}, {Gasparyan}, {Gaug}, {Giglietto},
  {Giordano}, {Gliwny}, {Godinovi{\'c}}, {Green}, {Hadasch}, {Hahn},
  {Heckmann}, {Herrera}, {Hoang}, {Hrupec}, {H{\"u}tten}, {Inada}, {Inoue},
  {Ishio}, {Iwamura}, {Jouvin}, {Kajiwara}, {Karjalainen}, {Kerszberg},
  {Kobayashi}, {Kubo}, {Kushida}, {Lamastra}, {Lelas}, {Leone}, {Lindfors},
  {Lombardi}, {Longo}, {L{\'o}pez}, {L{\'o}pez-Coto}, {L{\'o}pez-Oramas},
  {Loporchio}, {Machado de Oliveira Fraga}, {Maggio}, {Majumdar}, {Makariev},
  {Mallamaci}, {Maneva}, {Manganaro}, {Mannheim}, {Maraschi}, {Mariotti},
  {Mart{\'\i}nez}, {Mazin}, {Mender}, {Mi{\'c}anovi{\'c}}, {Miceli}, {Miener},
  {Minev}, {Miranda}, {Mirzoyan}, {Molina}, {Moralejo}, {Morcuende}, {Moreno},
  {Moretti}, {Munar-Adrover}, {Neustroev}, {Nigro}, {Nilsson}, {Ninci},
  {Nishijima}, {Noda}, {Nozaki}, {Ohtani}, {Oka}, {Otero-Santos}, {Palatiello},
  {Paneque}, {Paoletti}, {Paredes}, {Pavleti{\'c}}, {Pe{\~n}il}, {Perennes},
  {Persic}, {Prada Moroni}, {Prandini}, {Priyadarshi}, {Puljak}, {Rhode},
  {Rib{\'o}}, {Rico}, {Righi}, {Rugliancich}, {Saha}, {Sahakyan}, {Saito},
  {Sakurai}, {Satalecka}, {Schleicher}, {Schmidt}, {Schweizer}, {Sitarek},
  {{\v{S}}nidari{\'c}}, {Sobczynska}, {Spolon}, {Stamerra}, {Strom}, {Strzys},
  {Suda}, {Suri{\'c}}, {Takahashi}, {Tavecchio}, {Temnikov}, {Terzi{\'c}},
  {Teshima}, {Torres-Alb{\`a}}, {Tosti}, {Truzzi}, {van Scherpenberg}, {Vanzo},
  {Vazquez Acosta}, {Ventura}, {Verguilov}, {Vigorito}, {Vitale}, {Vovk},
  {Will}, {Zari{\'c}}, {Jorstad}, {Marscher}, {Boccardi}, {Casadio}, {Hodgson},
  {Kim}, {Krichbaum}, {L{\"a}hteenm{\"a}ki}, {Tornikoski}, {Traianou}, \&
  {Weaver}}]{hess+21}
{H.E.S.S. Collaboration}, {Abdalla}, H., {Adam}, R., {et~al.} 2021, \aap, 648,
  A23, \dodoi{10.1051/0004-6361/202038949}

\bibitem[{{HI4PI Collaboration} {et~al.}(2016){HI4PI Collaboration}, {Ben
  Bekhti}, {Fl{\"o}er}, {Keller}, {Kerp}, {Lenz}, {Winkel}, {Bailin},
  {Calabretta}, {Dedes}, {Ford}, {Gibson}, {Haud}, {Janowiecki}, {Kalberla},
  {Lockman}, {McClure-Griffiths}, {Murphy}, {Nakanishi}, {Pisano}, \&
  {Staveley-Smith}}]{HI4PI}
{HI4PI Collaboration}, {Ben Bekhti}, N., {Fl{\"o}er}, L., {et~al.} 2016, \aap,
  594, A116, \dodoi{10.1051/0004-6361/20162917810.48550/arXiv.1610.06175}

\bibitem[{{Holler} {et~al.}(2020){Holler}, {Lenain}, {de Naurois}, {Rauth}, \&
  {Sanchez}}]{holler20}
{Holler}, M., {Lenain}, J.~P., {de Naurois}, M., {Rauth}, R., \& {Sanchez},
  D.~A. 2020, Astroparticle Physics, 123, 102491,
  \dodoi{10.1016/j.astropartphys.2020.102491}

\bibitem[{{Isler} {et~al.}(2015){Isler}, {Urry}, {Bailyn}, {Smith}, {Coppi},
  {Brady}, {MacPherson}, {Hasan}, \& {Buxton}}]{Isleretal2015}
{Isler}, J.~C., {Urry}, C.~M., {Bailyn}, C., {et~al.} 2015, \apj, 804, 7,
  \dodoi{10.1088/0004-637X/804/1/7}

\bibitem[{{Kobulnicky} {et~al.}(2003){Kobulnicky}, {Nordsieck}, {Burgh},
  {Smith}, {Percival}, {Williams}, \& {O'Donoghue}}]{2003SPIE.4841.1634K}
{Kobulnicky}, H.~A., {Nordsieck}, K.~H., {Burgh}, E.~B., {et~al.} 2003, in
  Society of Photo-Optical Instrumentation Engineers (SPIE) Conference Series,
  Vol. 4841, Instrument Design and Performance for Optical/Infrared
  Ground-based Telescopes, ed. M.~{Iye} \& A.~F.~M. {Moorwood}, 1634--1644,
  \dodoi{10.1117/12.460315}

\bibitem[{{Lenain}(2018)}]{2018A+C....22....9L}
{Lenain}, J.~P. 2018, Astronomy and Computing, 22, 9,
  \dodoi{10.1016/j.ascom.2017.11.002}

\bibitem[{{MAGIC Collaboration} {et~al.}(2018){MAGIC Collaboration}, {Acciari},
  {Ansoldi}, {Antonelli}, {Arbet Engels}, {Arcaro}, {Baack}, {Babi{\'c}},
  {Banerjee}, {Bangale}, {Barres de Almeida}, {Barrio}, {Bednarek},
  {Bernardini}, {Berti}, {Besenrieder}, {Bhattacharyya}, {Bigongiari},
  {Biland}, {Blanch}, {Bonnoli}, {Carosi}, {Ceribella}, {Cikota}, {Colak},
  {Colin}, {Colombo}, {Contreras}, {Cortina}, {Covino}, {D'Elia}, {da Vela},
  {Dazzi}, {de Angelis}, {de Lotto}, {Delfino}, {Delgado}, {di Pierro}, {Do
  Souto Espi{\~n}era}, {Dom{\'\i}nguez}, {Dominis Prester}, {Dorner}, {Doro},
  {Einecke}, {Elsaesser}, {Fallah Ramazani}, {Fattorini},
  {Fern{\'a}ndez-Barral}, {Ferrara}, {Fidalgo}, {Foffano}, {Fonseca}, {Font},
  {Fruck}, {Galindo}, {Gallozzi}, {Garc{\'\i}a L{\'o}pez}, {Garczarczyk},
  {Gaug}, {Giammaria}, {Godinovi{\'c}}, {Guberman}, {Hadasch}, {Hahn},
  {Hassan}, {Herrera}, {Hoang}, {Hrupec}, {Inoue}, {Ishio}, {Iwamura}, {Kubo},
  {Kushida}, {Kuve{\v{z}}di{\'c}}, {Lamastra}, {Lelas}, {Leone}, {Lindfors},
  {Lombardi}, {Longo}, {L{\'o}pez}, {L{\'o}pez-Oramas}, {Maggio}, {Majumdar},
  {Makariev}, {Maneva}, {Manganaro}, {Mannheim}, {Maraschi}, {Mariotti},
  {Mart{\'\i}nez}, {Masuda}, {Mazin}, {Minev}, {Miranda}, {Mirzoyan}, {Molina},
  {Moralejo}, {Moreno}, {Moretti}, {Munar-Adrover}, {Neustroev}, {Niedzwiecki},
  {Nievas Rosillo}, {Nigro}, {Nilsson}, {Ninci}, {Nishijima}, {Noda},
  {Nogu{\'e}s}, {Paiano}, {Palacio}, {Paneque}, {Paoletti}, {Paredes},
  {Pedaletti}, {Pe{\~n}il}, {Peresano}, {Persic}, {Prada Moroni}, {Prandini},
  {Puljak}, {Garcia}, {Rhode}, {Rib{\'o}}, {Rico}, {Righi}, {Rugliancich},
  {Saha}, {Saito}, {Satalecka}, {Schweizer}, {Sitarek}, {{\v{S}}nidari{\'c}},
  {Sobczynska}, {Somero}, {Stamerra}, {Strzys}, {Suri{\'c}}, {Tavecchio},
  {Temnikov}, {Terzi{\'c}}, {Teshima}, {Torres-Alb{\`a}}, {Tsujimoto}, {van
  Scherpenberg}, {Vanzo}, {Vazquez Acosta}, {Vovk}, {Ward}, {Will},
  {Zari{\'c}}, {Fermi-Lat Collaboration}, {Becerra Gonz{\'a}lez}, {Raiteri},
  {Sandrinelli}, {Hovatta}, {Kiehlmann}, {Max-Moerbeck}, {Tornikoski},
  {L{\"a}hteenm{\"a}ki}, {Tammi}, {Ramakrishnan}, {Thum}, {Agudo}, {Molina},
  {G{\'o}mez}, {Fuentes}, {Casadio}, {Traianou}, {Myserlis}, \&
  {Kim}}]{magic18}
{MAGIC Collaboration}, {Acciari}, V.~A., {Ansoldi}, S., {et~al.} 2018, \aap,
  619, A159, \dodoi{10.1051/0004-6361/201833618}

\bibitem[{{Malkan} \& {Moore}(1986)}]{Malkanetal1986}
{Malkan}, M.~A., \& {Moore}, R.~L. 1986, \apj, 300, 216, \dodoi{10.1086/163796}

\bibitem[{{Mattox} {et~al.}(1996){Mattox}, {Bertsch}, {Chiang}, {Dingus},
  {Digel}, {Esposito}, {Fierro}, {Hartman}, {Hunter}, {Kanbach}, {Kniffen},
  {Lin}, {Macomb}, {Mayer-Hasselwander}, {Michelson}, {von Montigny},
  {Mukherjee}, {Nolan}, {Ramanamurthy}, {Schneid}, {Sreekumar}, {Thompson}, \&
  {Willis}}]{mattox96}
{Mattox}, J.~R., {Bertsch}, D.~L., {Chiang}, J., {et~al.} 1996, \apj, 461, 396,
  \dodoi{10.1086/177068}

\bibitem[{{Meyer} {et~al.}(2019){Meyer}, {Iyer}, {Reddy}, {Georganopoulos},
  {Breiding}, \& {Keenan}}]{Meyer+19}
{Meyer}, E.~T., {Iyer}, A.~R., {Reddy}, K., {et~al.} 2019, \apjl, 883, L2,
  \dodoi{10.3847/2041-8213/ab3db3}

\bibitem[{{Nalewajko} {et~al.}(2012){Nalewajko}, {Sikora}, {Madejski}, {Exter},
  {Szostek}, {Szczerba}, {Kidger}, \& {Lorente}}]{nalewajko+12}
{Nalewajko}, K., {Sikora}, M., {Madejski}, G.~M., {et~al.} 2012, \apj, 760, 69,
  \dodoi{10.1088/0004-637X/760/1/69}

\bibitem[{{Paliya} {et~al.}(2018){Paliya}, {Zhang}, {B{\"o}ttcher}, {Ajello},
  {Dom{\'\i}nguez}, {Joshi}, {Hartmann}, \& {Stalin}}]{Paliyaetal2018}
{Paliya}, V.~S., {Zhang}, H., {B{\"o}ttcher}, M., {et~al.} 2018, \apj, 863, 98,
  \dodoi{10.3847/1538-4357/aad1f0}

\bibitem[{{Parsons} \& {Hinton}(2014)}]{parsonshinton14}
{Parsons}, R.~D., \& {Hinton}, J.~A. 2014, Astroparticle Physics, 56, 26,
  \dodoi{10.1016/j.astropartphys.2014.03.002}

\bibitem[{{Phillips}(1978)}]{Phillips1978}
{Phillips}, M.~M. 1978, \apjs, 38, 187, \dodoi{10.1086/190553}

\bibitem[{{Poole} {et~al.}(2008){Poole}, {Breeveld}, {Page}, {Landsman},
  {Holland}, {Roming}, {Kuin}, {Brown}, {Gronwall}, {Hunsberger}, {Koch},
  {Mason}, {Schady}, {vanden Berk}, {Blustin}, {Boyd}, {Broos}, {Carter},
  {Chester}, {Cucchiara}, {Hancock}, {Huckle}, {Immler}, {Ivanushkina},
  {Kennedy}, {Marshall}, {Morgan}, {Pandey}, {de Pasquale}, {Smith}, \&
  {Still}}]{Poole08}
{Poole}, T.~S., {Breeveld}, A.~A., {Page}, M.~J., {et~al.} 2008, \mnras, 383,
  627, \dodoi{10.1111/j.1365-2966.2007.12563.x}

\bibitem[{{Prince} {et~al.}(2019){Prince}, {Gupta}, \& {Nalewajko}}]{prince+19}
{Prince}, R., {Gupta}, N., \& {Nalewajko}, K. 2019, \apj, 883, 137,
  \dodoi{10.3847/1538-4357/ab3afa}

\bibitem[{{Rakshit}(2020)}]{Rakshit2020}
{Rakshit}, S. 2020, A\&A, 642, A59, \dodoi{10.1051/0004-6361/202038324}

\bibitem[{{Reimer}(2007)}]{Reimer07}
{Reimer}, A. 2007, \apj, 665, 1023, \dodoi{10.1086/519766}

\bibitem[{{Roming} {et~al.}(2005){Roming}, {Kennedy}, {Mason}, {Nousek}, {Ahr},
  {Bingham}, {Broos}, {Carter}, {Hancock}, {Huckle}, {Hunsberger}, {Kawakami},
  {Killough}, {Koch}, {McLelland}, {Smith}, {Smith}, {Soto}, {Boyd},
  {Breeveld}, {Holland}, {Ivanushkina}, {Pryzby}, {Still}, \&
  {Stock}}]{Roming05}
{Roming}, P. W.~A., {Kennedy}, T.~E., {Mason}, K.~O., {et~al.} 2005, \ssr, 120,
  95, \dodoi{10.1007/s11214-005-5095-4}

\bibitem[{{Rybicki} \& {Lightman}(1979)}]{1979rpa..book.....R}
{Rybicki}, G.~B., \& {Lightman}, A.~P. 1979, {Radiative processes in
  astrophysics} (New York: Wiley-Interscience)

\bibitem[{{Saito} {et~al.}(2015){Saito}, {Stawarz}, {Tanaka}, {Takahashi},
  {Sikora}, \& {Moderski}}]{saito+15}
{Saito}, S., {Stawarz}, {\L}., {Tanaka}, Y.~T., {et~al.} 2015, \apj, 809, 171,
  \dodoi{10.1088/0004-637X/809/2/171}

\bibitem[{{Schlafly} \& {Finkbeiner}(2011)}]{schlaflyfinkbeiner11}
{Schlafly}, E.~F., \& {Finkbeiner}, D.~P. 2011, \apj, 737, 103,
  \dodoi{10.1088/0004-637X/737/2/103}

\bibitem[{Schutte {et~al.}(2022)Schutte, Britto, Böttcher, van Soelen, Marais,
  Kaur, Falcone, Buckley, Rajoelimanana, \& Cooper}]{Schutte_2022}
Schutte, H.~M., Britto, R.~J., Böttcher, M., {et~al.} 2022, The Astrophysical
  Journal, 925, 139, \dodoi{10.3847/1538-4357/ac3cb5}

\bibitem[{{Shakura} \& {Sunyaev}(1973)}]{shakurasunyaev73}
{Shakura}, N.~I., \& {Sunyaev}, R.~A. 1973, \aap, 500, 33

\bibitem[{{Zacharias} {et~al.}(2019){Zacharias}, {Dominis Prester},
  {Jankowsky}, {Lindfors}, {Meyer}, {Mohamed}, {Prokoph}, {Sanchez}, {Sitarek},
  {Terzic}, {Wagner}, {Wierzcholska}, {H.~E.~S.~S. Collaboration}, \& {MAGIC
  Collaboration}}]{zacharias+19}
{Zacharias}, M., {Dominis Prester}, D., {Jankowsky}, F., {et~al.} 2019,
  Galaxies, 7, 41, \dodoi{10.3390/galaxies7010041}

\end{thebibliography}
%
%
%
\appendix
\section{Supplementary Optical-UV SED and Spectropolarimetry Modeling} \label{sec:app1}
\begin{figure*}
\includegraphics[width=0.90\textwidth]{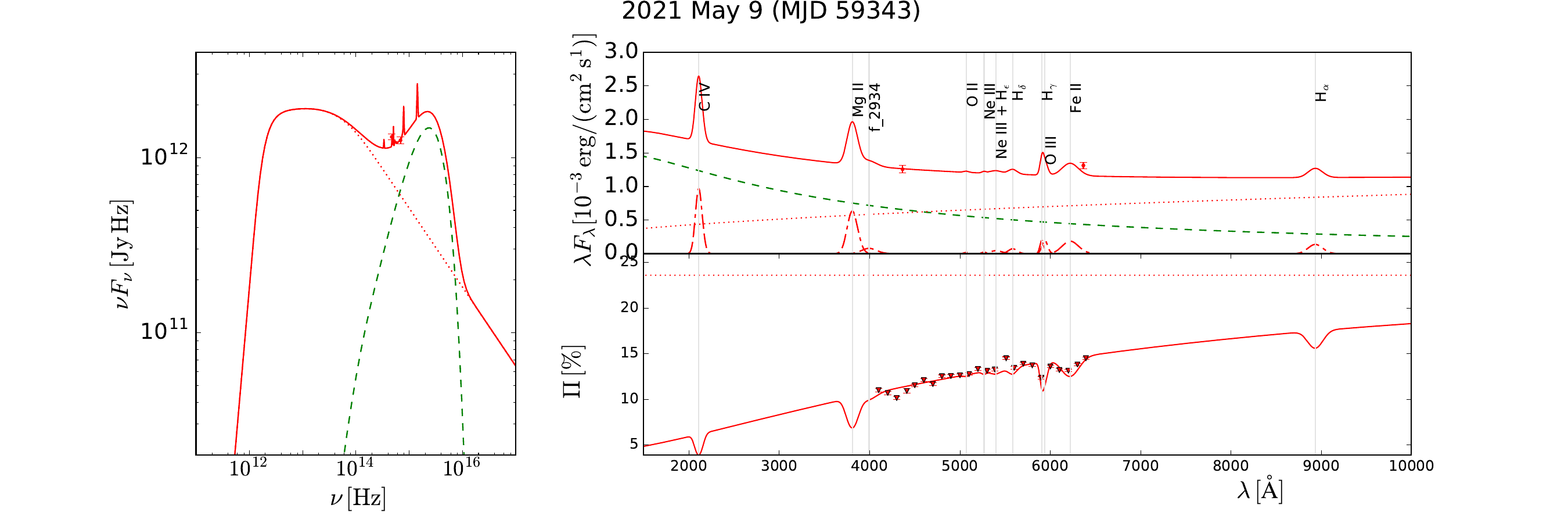}\\ \vspace*{-0.4cm}
\includegraphics[width=0.90\textwidth]{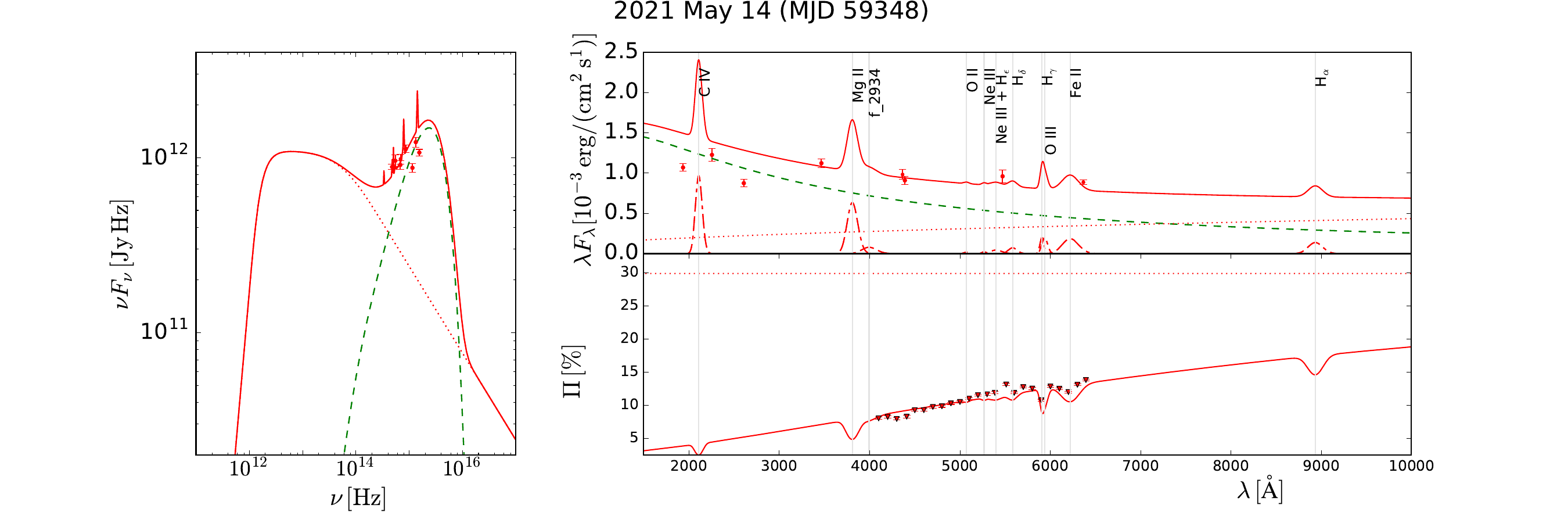}\\ \vspace*{-0.4cm}
\includegraphics[width=0.90\textwidth]{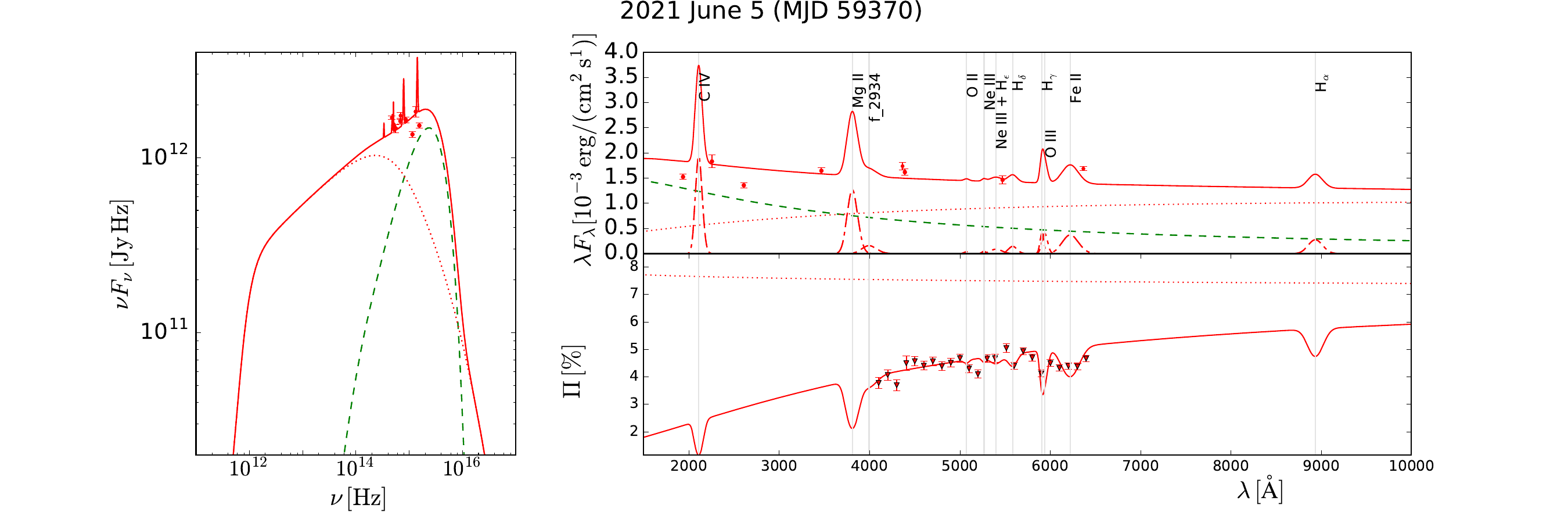}\\ \vspace*{-0.4cm}
\includegraphics[width=0.90\textwidth]{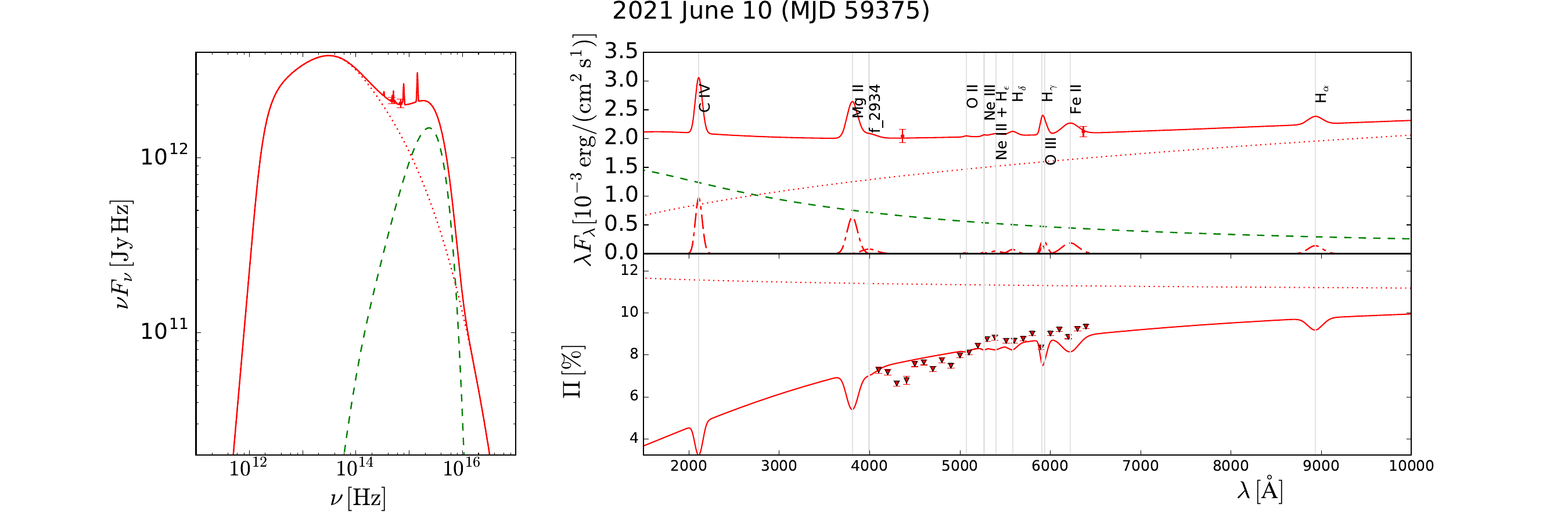}
\caption{Same as Fig.~\ref{fig:LEAverage} but for the individual observations in 2021.
}\label{fig:LE2021}
\end{figure*}

The plots in Fig. \ref{fig:LE2021} show the model fits to the optical-UV photometry and spectropolarimetry data for each of the SALT spectropolarimetry observing windows in 2021. Contemporaneous observations from the ATOM and \textit{Swift}-UVOT telescopes were included in the fits, when detections were obtained on the same day as the SALT detections except for the SALT observations of 2021 April 6 (MJD~59310), 2021 May 9 (MJD~59343) and 2021 June 10 (MJD~59375) where the ATOM data of 2021 April 7 (MJD~59311), 2021 May 8 (MJD~59342) and 2021 June 9 (MJD~59374) in the R-band were included, respectively, as guide to the fits. The parameters obtained with the model fit are given in Tab.~\ref{tab:OptUVpars2021} and the obtained line fluxes for each observation are listed in Tab.~\ref{tab:emlines}. The full spectropolarimetry results for each of the SALT observations are given in Tab.~\ref{tab:PolData_landscape}.

\begin{figure*}
\includegraphics[width=0.90\textwidth]{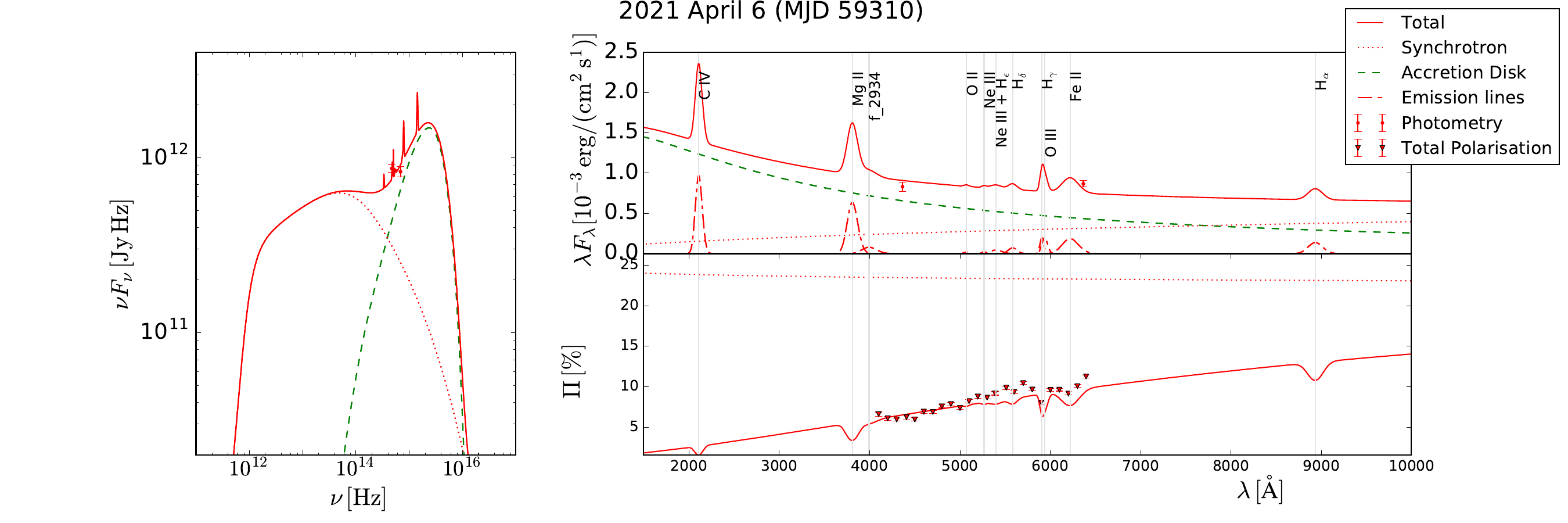}\\ \vspace*{-0.4cm}
\includegraphics[width=0.90\textwidth]{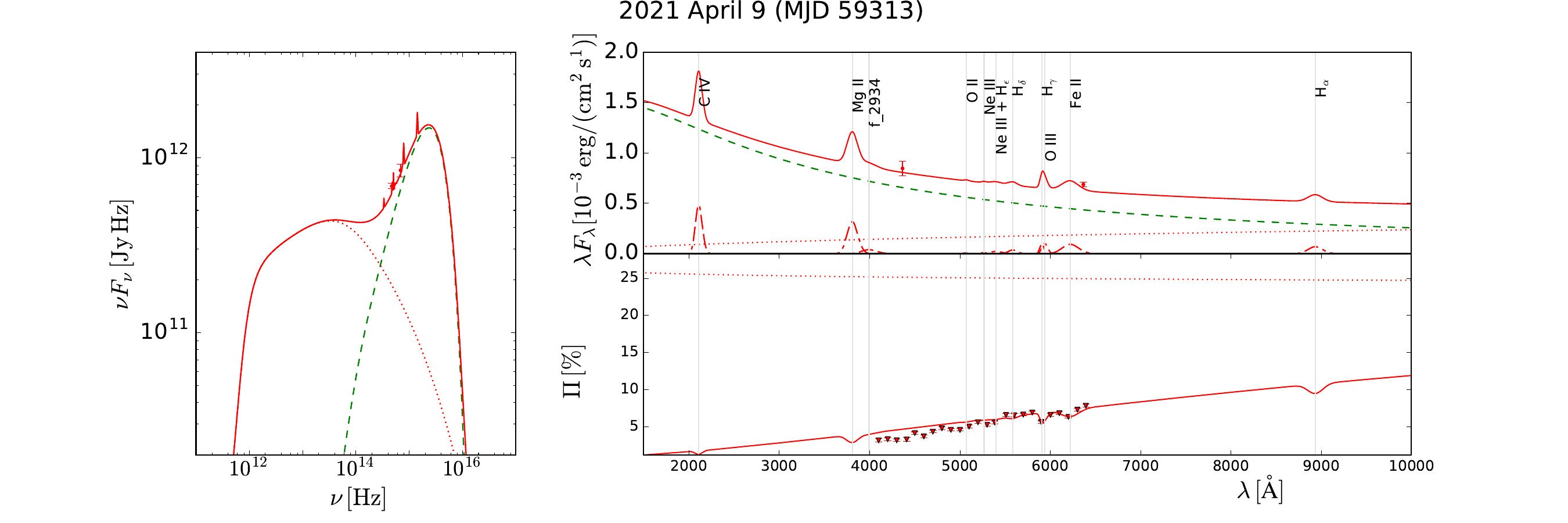}\\ \vspace*{-0.4cm}
\includegraphics[width=0.90\textwidth]{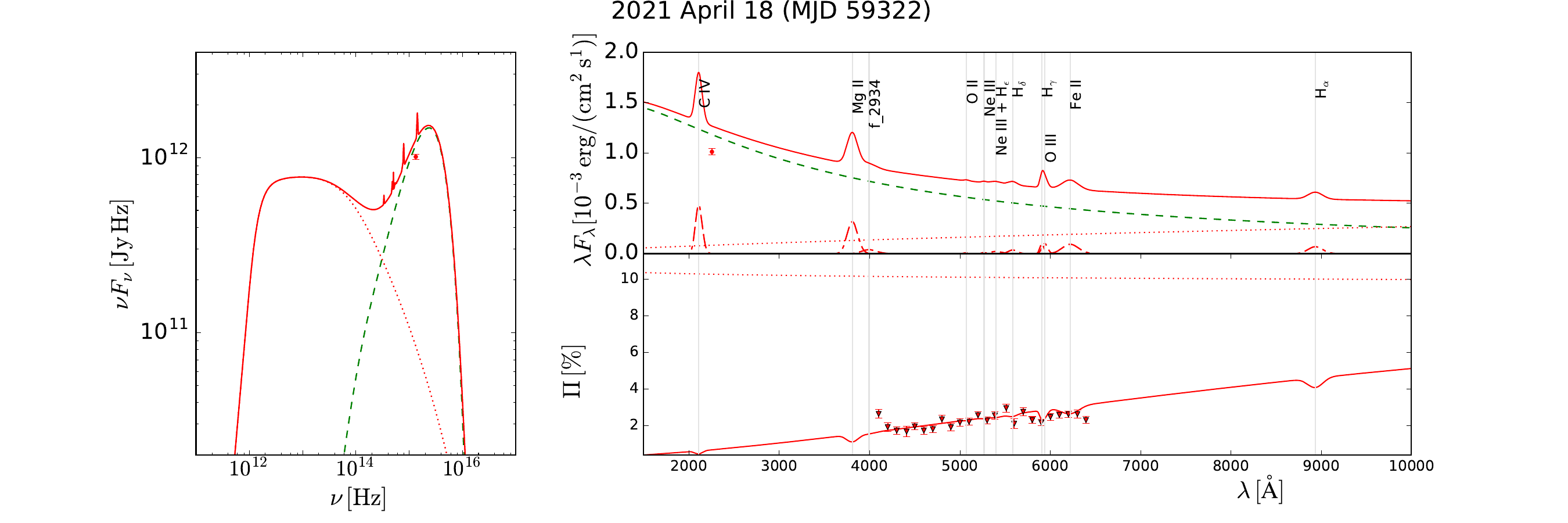}\\ \vspace*{-0.4cm}
\includegraphics[width=0.90\textwidth]{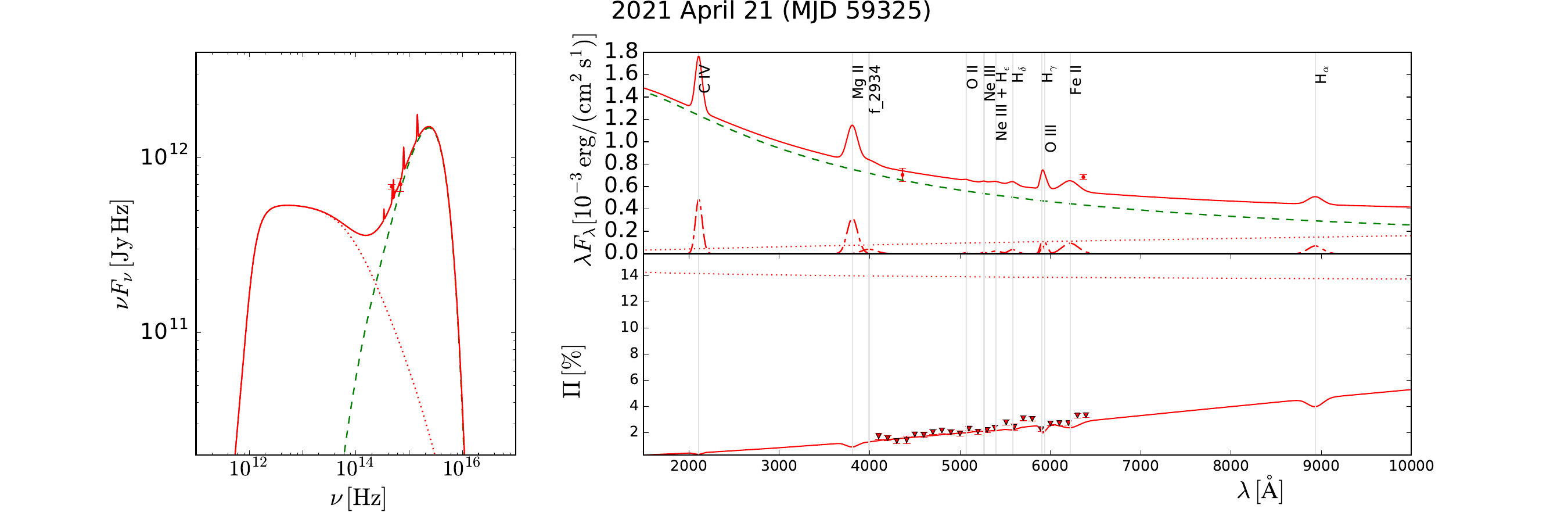}\\ 
\end{figure*}

\begin{table*}[!ht]
    \centering
    \caption{Parameters obtained by fitting the data from each of the 2021 observation windows with the code of \cite{Schutte_2022}. See Tab.~\ref{tab:OptUVpars} for information on the parameters that are taken as constant in the model throughout all states. For the goodness of fit $\chi^2_{pol}/ndf $, $ndf = 13$ for all states.}\label{tab:OptUVpars2021}
    \begin{tabular}{l cccccccc}
    \hline
        Date & MJD & $n_0$ & $\gamma_b$ & $\gamma_c$ &  $p_1$ &  $p_2$ & $F_B$ & $\chi^2_{pol}/ndf $ \\  \hline
       2021 April 06 & 59310 & $3.8 \times 10^{46}$ & $697$ & $ 2.8 \times 10^{3}$ & 2.5 & 3.5 & 0.28 & 0.25 \\ 
        2021 April 09 & 59313 & $4.6 \times 10^{46}$& 569 & $2.8 \times 10^{3}$ & 2.5 & 3.5 & 0.30 & 0.21 \\ 
        2021 April 18 & 59322 & $7.2 \times 10^{46}$ & 569 & $2.8 \times 10^{3}$ & 2.9 & 3.9 & 0.12 & 0.10 \\ 
        2021 April 21 & 59325 & $4.6 \times 10^{46}$ & 569 & $2.8 \times 10^{3}$ & 2.5 & 4.0 & 0.17 & 0.10 \\ 
        2021 May 09 & 59343 & $1.4 \times 10^{47}$ & 569 & $5.0 \times 10^{6}$ & 2.9 & 3.9 & 0.30 & 0.04 \\ 
        2021 May 14 & 59348 & $7.7 \times 10^{46}$ & 569 & $5.0 \times 10^{6}$ & 2.0 & 4.0 & 0.38 & 0.15 \\ 
        2021 June 05 & 59370 & $7.9\times 10^{45}$ & 156 & $2.8 \times 10^{3}$ & 2.4 & 3.4 & 0.09 & 0.03 \\ 
        2021 June 10 & 59375 & $6.0 \times 10^{47}$ & $493$ & $2.8 \times 10^{3}$ & 2.4 & 3.4 & 0.14 & 0.06 \\ 
       \hline 
    \end{tabular}
\end{table*}

The photometry fluxes decrease from 2021 April 6 to April 21 (MJD 59310 - 59325, excl. 2021 April 18, MJD 59322, for which ATOM data was not available, but a single very low \textit{Swift}-UVOT data point was recorded). On 2021 May 9 (MJD 59343), there is a sudden increase in flux, decreasing again on 14 May 2021, and thereafter, the flux continued increasing until 2021 June 10 (MJD 59375). 
The photometry fluxes and degree of polarization decreased/increased alongside each other, as shown in Fig.~\ref{fig:mwl_lightcurve}. 

The ordering of the magnetic fields, does not indicate the presence of a shock; in a shock-in-jet scenario, one expects that the ordering of the magnetic field decreases/increases in correlation with the degree of polarization and flux \citep{Paliyaetal2018}. Instead, the evolution of the ordering of the magnetic field shows no such correlation, which suggests the presence of turbulence and/or magnetic reconnection in the emission region as driver for the optical/UV variability. 

The $\chi^2_{pol}/ndf$ is the goodness of fit of the model to the spectropolarimetry data, where the number of degrees of freedom, ndf, is the amount of spectropolarimetry data points minus the amount of estimated parameters (equal to 10 in this model) minus 1. The goodness of the model fit to the few photometry data points is neglected and only applied to the abundent spectropolarimetry data, since fitting the prediction of the model's total polarization degree to the spectropolarimetry data is already dependent on the modeled total flux as well (where the modeled total flux was fitted to the photometry data). It does not indicate a good fit for all states since there might be contributions from components (such as emission lines) to the total flux (and thereby the total degree of polarization) that are missing or insufficiently accurately modeled. 
On the other hand, the inclusion of additional radiation components increases the number of free parameters in the model and therefore reduces its predictive power. Therefore, such additional components are not included. 



The \textit{Swift}-UVOT data shows an unexpected trend of a variable profile for each state. This might be explained by prominent emission lines that have fluxes higher than the continuum. 

In the spectropolarimetry data, the dominant line was identified as H$_\gamma$ from which the other lines were calculated relative to each other according to \cite{Phillips1978}. The remaining wavelength ranges of the emission lines are taken from \cite{Francisetal1991}, when available. The wavelength ranges of H$_\delta$, f$_{2934}$ and \mbox{$[$Ne III$]$ + H$_\epsilon$} (at $3967\,$\AA ) lines that are not given in \cite{Francisetal1991} are estimated by eye to fit the 
photometric and spectropolarimetric data. 


\begin{table*}
 \centering
    \caption{Fluxes of the emission lines of \source\ during SALT spectropolarimetry observations in 2021. }\label{tab:emlines}
    \begin{tabular}{c|c|c|c|c|c}
     \hline
        & Dates & 2021 April 6  & 2021 April 9, 21 & 2021 April 18 & 2021 June 05  \\ 
        & & 2021 May 9,14 &   & & \\ 
        & & 2021 June 10 & & &  \\ \hline
    
        & MJD & 59310     & 59313, 59325 & 59322  & 59370  \\ 
        & &  59343, 59348 &  & & \\ 
        & &  59375 & & &\\ \hline
        
    Emission   & Restframe & \multicolumn{4}{c}{Line flux}    \\
    line & wavelength (\AA) & \multicolumn{4}{c}{($\times 10^{12} $ Jy Hz)} \\ \hline
        H$\alpha$ & 6563 & 0.14 & 0.07 & 0.07 & 0.28\\
        Fe II & 4570 & 0.19 & 0.09 & 0.1 & 0.37   \\ 
        $[$O III$]$ & 4363 & 0.2 & 0.1 & 0.1 &0.4   \\ 
        H$_{\gamma}$ & 4340 & 0.2 & 0.1 & 0.1 & 0.4   \\ 
        H$_{\delta}$ & 4102 & 0.07 & 0.04 & 0.03 & 0.14   \\ 
        $[$Ne III$]$ + H$_\epsilon$ & 3967 & 0.04 & 0.02  & 0.02 & 0.09   \\ 
        $[$Ne III$]$  & 3869 &  0.02 & 0.01  & 0.01 & 0.04  \\ 
        $[$O II$]$ & 3727 & 0.01 & 0.01 & 0.01 & 0.04  \\ 
        $f_{2934}$ & 2934 & 0.08 & 0.03 & 0.04 & 1.15  \\ 
        Mg II & 2798 & 0.63 & 0.32 & 0.31 & 1.27  \\ 
        C IV & 1549& 0.97 & 0.48 & 0.48 &1.94  \\ 
        \hline
    \end{tabular}
\end{table*}

\movetabledown=50mm
\begin{table*}
\begin{rotatetable*}
\begin{center}
\caption{The average degree of linear polarization and average equatorial polarization angle for each of the SALT observations, in four different wavelength ranges. The wavelength ranges (as specified in Section 2.2) correspond to $\lambda_{\rm range 1} = 3670 - 4060\,$\AA, $\lambda_{\rm range 2} = 4100 - 4400\,$\AA, $\lambda_{\rm range 3} = 4480 - 4780\,$\AA, and $\lambda_{\rm range 4} = 4800 - 5100\,$\AA.} \label{tab:PolData_landscape}
\begin{tabular}{l|cccc|cccc}
    \hline
     & \multicolumn{4}{c|}{Degree of Linear Polarization (\%)} & \multicolumn{4}{c}{Equatorial Polarization Angle ($^{\circ}$)} \\ \hline
        Date & $\langle \Pi \rangle_{\rm range 1}$ & $\langle \Pi \rangle_{\rm range 2}$ & $\langle \Pi \rangle_{\rm range 3}$ &  $\langle \Pi \rangle_{\rm range 4}$ & $\langle \rm PA \rangle_{\rm range 1}$ & $\langle \rm PA \rangle_{\rm range 2}$ & $\langle \rm PA \rangle_{\rm range 3}$ &  $\langle \rm PA \rangle_{\rm range 4}$ \\  \hline
        2021 April 06 & 5.01 $\pm$ 0.76 & 6.08 $\pm$ 0.24 & 6.65 $\pm$ 0.35 & 7.75 $\pm$ 0.29 & 20.65 $\pm$ 2.89 & 16.54 $\pm$ 0.58 & 13.84 $\pm$ 1.02 & 12.91 $\pm$ 0.92\\
        2021 April 09 & 2.31 $\pm$ 0.60 & 3.24 $\pm$ 0.23 & 4.10 $\pm$ 0.21 & 4.70 $\pm$ 0.28 & 1.16 $\pm$ 2.85 & 8.94 $\pm$ 1.92 & 6.40 $\pm$ 1.51 & 2.95 $\pm$ 1.10 \\
        2021 April 18 & 1.93 $\pm$ 0.46 & 1.93 $\pm$ 0.25 & 1.72 $\pm$ 0.16 & 2.16 $\pm$ 0.20 & 106.48 $\pm$ 7.39 & 100.99 $\pm$ 1.79 & 101.29 $\pm$ 2.38 & 97.63 $\pm$ 6.95 \\
        2021 April 21 & 1.21 $\pm$ 0.38 & 1.58 $\pm$ 0.45 & 1.87 $\pm$ 0.23 & 2.10 $\pm$ 0.18 & 172.81 $\pm$ 13.22 & 173.28 $\pm$ 5.71 & 180.59	$\pm$ 1.24 & 181.37 $\pm$ 5.23 \\
        2021 May 09   & 9.06 $\pm$ 1.16 & 10.46 $\pm$ 0.45 & 11.84 $\pm$ 0.36 & 12.62 $\pm$ 0.16 & 153.70 $\pm$ 1.64 & 155.66 $\pm$ 0.56 & 155.84 $\pm$ 0.46 & 155.20 $\pm$ 0.81 \\
        2021 May 14   & 6.45 $\pm$ 1.51 & 8.07 $\pm$ 0.25 & 9.55 $\pm$ 0.41 & 10.50 $\pm$ 0.28 & 33.61 $\pm$ 1.65 & 33.17 $\pm$ 0.92  & 32.33 $\pm$ 0.88 & 31.77 $\pm$ 0.39 \\
        2021 June 05  & 3.23 $\pm$ 0.56 & 4.06 $\pm$ 0.41 & 4.45 $\pm$ 0.32 & 4.50 $\pm$ 0.11 & 150.68 $\pm$ 1.75 & 145.80 $\pm$ 1.49 & 144.92 $\pm$ 1.19 & 145.48 $\pm$ 1.74 \\
        2021 June 10  & 6.07 $\pm$ 0.64 & 6.95 $\pm$ 0.40 & 7.48 $\pm$ 0.20 & 7.89 $\pm$ 0.25 & 80.21 $\pm$ 0.75 & 79.82 $\pm$ 0.99 & 80.40 $\pm$ 0.52 & 80.79 $\pm$ 0.79 \\
        2022 April 25 & 1.11 $\pm$ 0.44 & 1.17 $\pm$ 0.56 & 1.63 $\pm$ 0.97 & 1.53 $\pm$ 0.28 & 78.89 $\pm$ 30.56 & 63.34 $\pm$ 13.10 & 67.06 $\pm$ 33.03 & 53.45 $\pm$ 24.00 \\
        2022 April 26 & 1.55 $\pm$ 1.05 & 1.49 $\pm$ 0.65 & 0.78 $\pm$ 0.37 & 1.02 $\pm$ 0.39 & 51.43 $\pm$ 15.22 & 60.89 $\pm$ 12.52 & 36.54 $\pm$ 15.11 & 65.31 $\pm$ 30.40 \\
        2022 May 24 & 1.60 $\pm$ 0.53 & 1.52 $\pm$ 0.61 & 1.34 $\pm$ 0.58 & 1.96 $\pm$ 0.56 & 62.99 $\pm$ 12.08 & 59.34 $\pm$ 13.59 & 60.64 $\pm$ 14.07 & 55.17 $\pm$ 10.72 \\
        2022 May 25 & 1.00 $\pm$ 0.62 & 1.01 $\pm$ 0.29 & 1.12 $\pm$ 0.39 & 1.01 $\pm$ 0.32 & 88.53 $\pm$ 37.74 & 81.16 $\pm$ 11.77 & 76.19 $\pm$ 12.20 & 76.41 $\pm$ 10.86 \\
        2022 May 30 & 1.41 $\pm$ 0.76 & 1.75 $\pm$ 1.08 & 2.25 $\pm$ 1.07 & 2.01 $\pm$ 0.85 & 66.91 $\pm$ 28.68 & 59.95 $\pm$ 10.04 & 48.02 $\pm$ 13.75 & 50.79 $\pm$ 16.49 \\
        2022 June 05 & 1.61 $\pm$ 0.65 & 1.32 $\pm$ 0.33 & 0.86 $\pm$ 0.29 & 1.19 $\pm$ 0.44 & 104.78 $\pm$ 18.25 & 83.35 $\pm$ 5.92 & 67.60 $\pm$ 14.73 & 66.62 $\pm$ 9.63 \\
        2022 June 20 & 1.91 $\pm$ 0.59 & 1.03 $\pm$ 0.41 & 1.66 $\pm$ 0.25 & 1.16 $\pm$ 0.24 & 35.88 $\pm$ 28.23 & 60.98 $\pm$ 17.17 & 73.67 $\pm$ 10.36 & 65.40 $\pm$ 16.59 \\
        2022 June 26 & 1.24 $\pm$ 0.48 & 0.61 $\pm$ 0.29 & 1.18 $\pm$ 0.35 & 1.15 $\pm$ 0.65 & 112.92 $\pm$ 33.00 & 92.35 $\pm$ 44.39 & 85.73 $\pm$ 9.38 & 90.41 $\pm$ 16.94 \\
        2022 July 27 & 1.11 $\pm$ 0.48 & 1.45 $\pm$ 0.42 & 1.17 $\pm$ 0.53 & 1.24 $\pm$ 0.31 & 70.68 $\pm$ 33.54 & 88.82 $\pm$ 17.95 & 48.30 $\pm$ 32.87 & 64.05 $\pm$ 15.81 \\
        2022 July 28 & 1.79 $\pm$ 1.29 & 2.29 $\pm$ 0.97 & 1.56 $\pm$ 0.77 & 1.29 $\pm$ 0.46 & 57.58$\pm$ 26.74 & 85.24 $\pm$ 9.63 & 67.96 $\pm$ 35.24 & 85.87 $\pm$ 19.82 \\
        2022 July 31 & 1.57 $\pm$ 0.75 & 1.60 $\pm$ 0.56 & 1.38 $\pm$ 0.31 & 1.13 $\pm$ 0.27 & 69.96 $\pm$ 2.17 & 60.70 $\pm$ 9.13 & 74.41 $\pm$ 13.54 & 54.01 $\pm$ 9.69 \\
       \hline
\end{tabular}
\end{center}
\end{rotatetable*}
\end{table*}

\section{X-ray spectral analysis} \label{app:xray}
\begin{table*}
    \centering
    \caption{Parameters of the power-law spectral fits to the \textit{Swift}-XRT 2021 and 2022 observations. The last two columns give the average flux above $0.3\,$keV and its p-value of compatibility with a constant.}\label{tab:xrtdata}
    \begin{tabular}{lccccc}
    \hline
    Year &  Time exposure  & Normalization  & Photon index & Avg. flux  & p-value \\
     & [ks] & [cm$^{-2}$\,s$^{-1}$\,keV$^{-1}$] & & [erg\,cm$^{-2}$\,s$^{-1}$] & \\
    \hline
    2021 & 8.6  & $(1.11 \pm 0.04) \times 10^{-3}$ &  $1.51 \pm 0.04$ & $(8.4 \pm 0.4)\E{-12}$ & 0.85 \\
    2022 & 15.3 & $(0.74 \pm 0.03) \times 10^{-3}$ &  $1.47 \pm 0.04$ & $(5.9 \pm 0.2)\E{-12}$ & 0.007 \\
    \hline
     \end{tabular}
\end{table*} 
Table~\ref{tab:xrtdata} provides an overview over the spectral results of the \textit{Swift}-XRT analysis for both years.

\end{document}